\documentclass[sigconf]{acmart}
\usepackage{multirow}
\usepackage{algorithm}
\usepackage[noend]{algpseudocode}
\usepackage{amsmath}
\usepackage{subcaption} 
\usepackage{pgfplots}
\usepackage[normalem]{ulem}
\usepackage[skip=5pt]{caption}
\usetikzlibrary{pgfplots.groupplots}
\pgfplotsset{width=4cm,compat=1.9}
\makeatletter
\def\BState{\State\hskip-\ALG@thistlm}
\makeatother

\newcommand{\true}{\text{T}}
\newcommand{\false}{\text{F}}
\newcommand{\mtrue}{\textbf{T}}
\newcommand{\mfalse}{\textbf{F}}

\AtBeginDocument{%
  \providecommand\BibTeX{{%
    \normalfont B\kern-0.5em{\scshape i\kern-0.25em b}\kern-0.8em\TeX}}}

\copyrightyear{2021}
\acmYear{2021} 
\setcopyright{iw3c2w3}
\acmConference[WWW '21]{Proceedings of the Web Conference 2021}{April 19--23, 2021}{Ljubljana, Slovenia} 
\acmBooktitle{Proceedings of the Web Conference 2021 (WWW '21), April 19--23, 2021, Ljubljana, Slovenia}
\acmPrice{}
\acmDOI{10.1145/3442381.3449973}
\acmISBN{978-1-4503-8312-7/21/04}

\begin{document}

\title{Neural Collaborative Reasoning}

 \author{Hanxiong Chen}
 \affiliation{%
  \institution{Rutgers University}
  \institution{New Brunswick, NJ, US}
    }
 \email{hanxiong.chen@rutgers.edu}

\author{Shaoyun Shi}
 \affiliation{%
  \institution{Tsinghua University}
  \institution{Beijing, China}
    }
 \email{shisy17@mails.tsinghua.edu.cn}
 
\author{Yunqi Li}
 \affiliation{%
  \institution{Rutgers University}
  \institution{New Brunswick, NJ, US}
    }
 \email{yunqi.li@rutgers.edu}
 
\author{Yongfeng Zhang}
 \affiliation{%
  \institution{Rutgers University}
  \institution{New Brunswick, NJ, US}
    }
 \email{yongfeng.zhang@rutgers.edu}

\begin{abstract}
Existing Collaborative Filtering (CF) methods are mostly designed based on the idea of matching, i.e., by learning user and item embeddings from data using shallow or deep models, they try to capture the associative relevance patterns in data, so that a user embedding can be matched with relevant item embeddings using designed or learned similarity functions. However, as a cognition rather than a perception intelligent task, recommendation requires not only the ability of pattern recognition and matching from data, but also the ability of cognitive reasoning in data.

In this paper, we propose to advance Collaborative Filtering (CF)\\ to Collaborative Reasoning (CR), which means that each user knows part of the reasoning space, and they collaborate for reasoning in the space to estimate preferences for each other. 
Technically, we propose a Neural Collaborative Reasoning (NCR) framework to bridge learning and reasoning. Specifically, we integrate the power of representation learning and logical reasoning, where representations capture similarity patterns in data from perceptual perspectives, and logic facilitates cognitive reasoning for informed decision making. An important challenge, however, is to bridge differentiable neural networks and symbolic reasoning in a shared architecture for optimization and inference. To solve the problem, we propose a modularized reasoning architecture, which learns logical operations such as AND ($\wedge$), OR ($\vee$) and NOT ($\neg$)
as neural modules for implication reasoning ($\rightarrow$).
In this way, logical expressions can be equivalently organized as neural networks, so that logical reasoning and prediction can be conducted in a continuous space. Experiments on real-world datasets verified the advantages of our framework compared with both shallow, deep and reasoning models.


\end{abstract}

%

%


%
\maketitle

\section{Introduction}

\begin{figure}[t]
\centering
\includegraphics[scale=0.33]{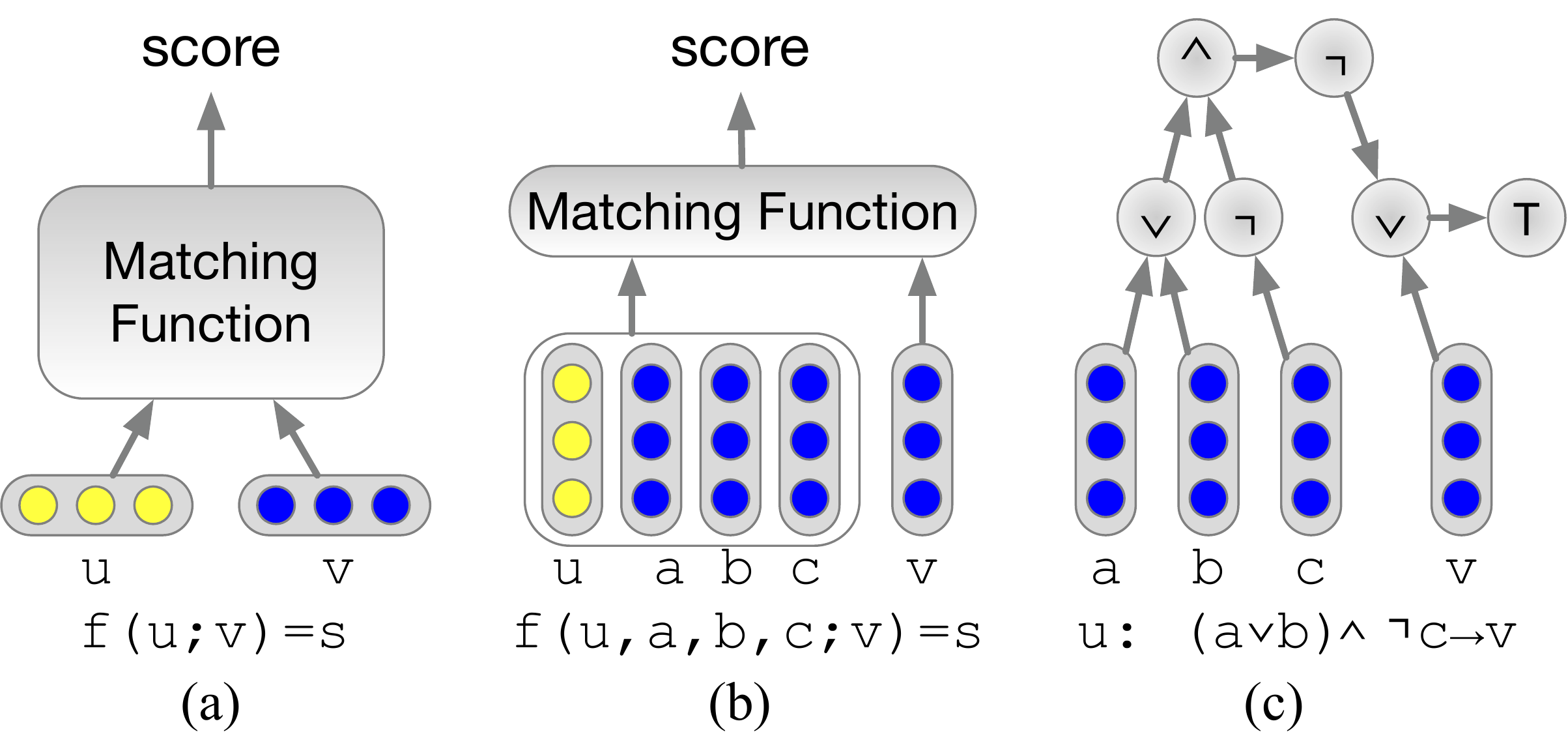}
\vspace{-5pt}
\caption{An overview of the fundamental structure of different collaborative filtering algorithms.}
\label{fig:overview}
\vspace{-15pt}
\end{figure}

Collaborative Filtering (CF) is an important approach to recommender systems \cite{ekstrand2011collaborative,rs}. By leveraging the wisdom of crowd, CF methods predict a user's future preferences based on his or her previous records. Many existing CF methods are designed based on the fundamental idea of similarity matching, with either designed or learned matching functions, as illustrated in Figure \ref{fig:overview}(a). For example, early CF algorithms, such as User-based CF \cite{resnick1994grouplens} and Item-based CF \cite{sarwar2001item}, consider the row and column vectors in the original user-item rating matrix as the user and item representations (i.e., embedding), and a manually designed weighted average function is used as the matching function $f(\cdot)$ to calculate the relevance score between each user $u$ and a candidate item $v$. The advance of machine learning has further extended CF methods for improved accuracy. One prominent example is Matrix Factorization (MF) techniques for CF \cite{koren2009matrix}, which takes inner product as the matching function $f(\cdot)$, and learns the user and item embeddings in the inner product space to fit ground-truth user-item interactions.

Researchers have further explored CF under the similarity matching framework.
One approach is to learn better embeddings. For example, context-aware CF integrates context information such as time and location to learn informative embeddings \cite{cf-temporal,multiverse,cars}, and heterogeneous information sources can be used to enrich the embeddings \cite{zhang2017joint}, such as text \cite{zheng2017joint}, image \cite{he2016vbpr}, and knowledge graphs \cite{zhang2016collaborativekdd,ai2018learning}. We can also explicitly consider a user's behavior history to learn better embeddings (Figure \ref{fig:overview}(b)), such as in sequential recommendation \cite{hidasi2016session,chen2018sequential,kang2018self,li2017neural}. Another approach is to learn better matching functions. For example, using vector translation instead of inner product for matching \cite{he2017translation}, or learning the matching function based on metric learning \cite{hsieh2017collaborative} and neural networks \cite{cheng2016wide,xue2017deep,he2017neural}. However, whether complex neural matching functions are better than simple matching functions is controversial \cite{dacrema2019we,rendle2020neural,ferrari2020critically,dacrema2021troubling}.



Similarity matching-based CF methods have been adopted in many real-world recommender systems. However, as a cognition rather than a perception task, recommendation requires not only the ability of pattern learning and matching, but also the ability of cognitive reasoning, because a user's future behavior may not be simply driven by its similarity with the user's previous behaviors, but instead by the user's cognitive reasoning procedure about what to do next. For example, if a user has purchased a laptop before, this does not lead to the user purchasing similar laptops in the future, rather, one would expect the user to purchase further equipment such as a laptop bag. Such a reasoning procedure may exhibit certain logical structures, such as $(a\vee b)\wedge\neg c\rightarrow v$, as shown in Figure \ref{fig:overview}(c), which means that if the user likes $a$ \textit{or} $b$, \textit{and} does \textit{not} like $c$, \textit{then} he/she would probability like $v$. 
In a broader sense, the community has realized the importance of advancing AI from perception to cognition tasks \cite{bengio2019from,marcus2020next,shi2020neural}. As a representative cognitive reasoning task, we hope an intelligent recommendation system would be able to conduct logical reasoning over the data to predict user's future behaviors for personalized recommendation.

To achieve this goal, we propose Neural Collaborative Reasoning (NCR), which defines the recommendation problem as a differentiable (i.e., neural) logical reasoning problem based on the wisdom of the crowd (i.e., collaborative). Specifically, each user's behavior record is considered as a Horn clause, such as $(a\vee b)\wedge\neg c\rightarrow v$ in the above example, meaning that the user liked item $v$ given his/her previous preferences on $a, b$ and $c$.
In this sense, each user contributes to part of the whole logical space, so that we can conduct collaborative logical reasoning based on the collective information from all users to estimate the preferences for each user. More specially, we propose a neural logic reasoning architecture,
which integrates the power of embedding learning and logical reasoning in a shared model. The model learns basic logical operations such as AND ($\wedge$), OR ($\vee$), and NOT ($\neg$) as neural modules based on logic regularization. As a result, the recommendation problem can be formalized as estimating the probability that a Horn clause is True (\true),
such as $(a\vee b)\wedge\neg c\rightarrow v$ in the example. Based on the definition of material implication ($\rightarrow$)\footnote{Material implication ($\rightarrow$) can be represented by basic operations: $x\rightarrow y \Leftrightarrow \neg x \vee y$}, this reduces to the T/F evaluation of $\neg ((a\vee b)\wedge\neg c) \vee v$, which only includes basic logical operations. Finally, the Horn clause can be identically transformed into a neural architecture using the logical neural modules, as shown in Figure \ref{fig:overview}(c), which decides the T/F value of the expression. In this way, differentiable neural networks and symbolic reasoning are bridged in a shared architecture for optimization and inference. 

The key contributions of the paper are as follows:
\begin{itemize}
    \item We propose a novel neural collaborative reasoning framework to bridge symbolic logical reasoning and continuous embedding learning for recommendation.
    \item We propose to adopt Horn clause for implication reasoning in recommendation, which naturally fits with the prediction nature of recommendation tasks.
    \item We propose a neural logic reasoning architecture, which dynamically constructs the network structure according to the given logical expression, and enables logic priors to be added to the neural network.
    \item We conduct experiments on several real-world recommendation datasets to analyze the behavior of our framework.
\end{itemize}


The following part of the paper will include related work (section \ref{sec:related}), preliminaries (section \ref{sec:preliminaries}), our framework (section \ref{sec:model}), experiments (section \ref{sec:experiment}), as well as conclusions and future work (section \ref{sec:conclusion}).


\section{Related Work}\label{sec:related}
Collaborative Filtering (CF) has been an important approach to recommender systems. Due to its long-time research history and the wide scope of literature, it is hardly possible to cover all CF algorithms, so we review some representative methods in this section, and a more comprehensive review can be seen in \cite{ekstrand2011collaborative,zhang2019deep,zhang2020explainable}.

Early approaches to CF consider the user-item rating matrix and conduct rating prediction with user-based \cite{resnick1994grouplens,konstan1997grouplens} or item-based \cite{sarwar2001item,Linden2003} collaborative filtering methods. With the development of dimension reduction methods, latent factor models such as matrix factorization are later widely adopted in recommender systems, such as singular value decomposition \cite{koren2009matrix}, non-negative matrix factorization \cite{lee2001algorithms}, and probabilistic matrix factorization \cite{mnih2008probabilistic}. In these approaches, each user and item is learned as a latent vector to calculate the matching score of the user-item pairs.

Recently, the development of deep learning and neural network models has further extended collaborative filtering methods for recommendation. The relevant methods can be broadly classified into two sub-categories: similarity learning approach, and representation learning approach. The similarity learning approach adopts simple user/item representations (such as one-hot) and learns a complex matching function (such as a prediction network) to calculate user-item matching scores \cite{xue2017deep,he2017neural,he2017translation,hsieh2017collaborative,cheng2016wide}, while the representation learning approach learns rich user/item representations and adopts a simple matching function (e.g., inner product) for efficient matching score calculation \cite{zheng2017joint,zhang2017joint,zhang2016collaborativekdd,ai2018learning,mcauley2015image}. However, there exist debates over whether complex matching functions are better than simple functions \cite{dacrema2019we,rendle2020neural,ferrari2020critically,dacrema2021troubling}.
Another important direction is learning to rank for recommendation, which learns the relative ordering of items instead of the absolute preference scores. A representative method is Bayesian personalized ranking (BPR) \cite{bpr}, which is a pair-wise learning to rank method. It is also further generalized to take other information sources such as images \cite{he2016vbpr}.

Although many CF approaches have been developed for recommendation tasks, existing methods mostly model recommendation as a perception task based on similarity matching instead of a cognition task based on cognitive/logical reasoning. However, users' future behaviors may not be simply driven by the similarity with their previous behavior, but a concrete reasoning procedure about what to do next.
Integrating logical reasoning and neural networks has been considered in several research contexts. 
According to \cite{besold2017neural}, connectionism in AI can date back to 1943 \cite{mccelloch1943logical}, which is arguably the first neural-symbolic system for Boolean logic. 
More recently, it is shown that argumentation frameworks, abductive reasoning, and normative multi-agent systems can also be represented by neural symbolic frameworks \cite{besold2017neural,garcez2012neural,dai2019bridging,dong2019neural,jiang2019neural}.
Another approach to integrating machine learning and logical reasoning is Markov logic networks \cite{richardson2006markov,qu2019probabilistic,zhang2020efficient}, which combines probabilistic graphical models with first-order logic. It leverages domain knowledge and logic rules to learn graph structure for inference, which is effective for reasoning on knowledge graphs \cite{qu2019probabilistic}.

The most related work to ours is neural logic reasoning \cite{shi2020neural}, which adopts neural logic modules for solving logical equations and (non-personalized) recommendation. However, our work is different on three aspects: we build neural models for logical reasoning based on the implication form of Horn clauses, which is a more natural way of making logical predictions in recommendation tasks; we develop a personalized recommendation model while the model in \cite{shi2020neural} can only conduct non-personalized recommendation; we explore if and how different neural logic structures influence the prediction performance of the neural logic models.

More broadly, researchers have realized the importance of advancing AI from perception to cognition tasks \cite{bengio2019from,marcus2020next}.
As a representative cognition task, we hope future intelligent recommendation systems can model higher-level cognitive intelligence for informed planning, reasoning, and decision making.

\section{Preliminaries}\label{sec:preliminaries}
In this section, we briefly introduce some logical operators and basic logical laws used in this work. We start from propositional logic, which includes three basic operations: AND (conjunction), OR (disjunction), and NOT (negation). Each variable, such as $x$, is called a $\textit{literal}$. A flat operation over literals, such as $(x\wedge y)$, is called a $\textit{clause}$, while operations over clauses, such as $(x\wedge y)\vee (a\wedge b\wedge c)$, is called an $\textit{expression}$. Each logical operation should satisfy some basic laws in propositional logic, for example, the double negation law of NOT: $\neg(\neg x)=x$. We list some laws used in our work in Table \ref{tb:regularizer}. Another useful law not listed in the table is the De Morgan's Law, which states that:
\begin{equation}
\label{morgan}
    \begin{split}
&\neg(x \wedge y) \Leftrightarrow \neg x \vee \neg y\\
&\neg(x \vee y) \Leftrightarrow \neg x \wedge \neg y 
    \end{split}
\end{equation}

Different from neural logic reasoning \cite{shi2020neural}, besides these operations and laws, we introduce another secondary logical operation $x\rightarrow y$ called material implication, which is fundamental to logic reasoning with Horn clauses, and as we will show later, it naturally fits into the prediction task of personalized recommendation. This operation can be equivalently transformed using basic operations:
\begin{equation}
\label{implication}
    x\rightarrow y \Leftrightarrow \neg x\vee y
\end{equation}

Propositional logic is a very useful language for symbolic reasoning. However, the symbolic nature of the language makes it difficult to be ``learned'' from data based on continuous optimization. To solve the problem, we borrow the idea of distributed representation learning \cite{mikolov2013distributed}, and propose a neural-symbolic framework for logical reasoning in a continuous space. Similar to \cite{shi2020neural}, each literal $x$ is learned as a vector embedding $\textbf{x}$, and each logical operation (e.g., $\wedge$) is learned as a neural module (e.g., $\textbf{z}=\text{AND}(\textbf{x},\textbf{y})$). As a result, an expression can be organized as a neural architecture (toy example in Figure \ref{fig:overview}(c), and more details later), which evaluates the T/F value of the expression in a latent space.


\section{Neural Collaborative Reasoning}\label{sec:model}
We present our Neural Collaborative Reasoning (NCR) framework in this section, which encapsulates logical reasoning into a dynamic neural architecture. 
We first formalize recommendation into a logical reasoning problem in Horn clause form. Then we introduce how to dynamically assemble the literals and logical operations into a neural network for recommendation. After that, we present the logical regularizers, which regularize the behavior of neural modules to conduct the expected logical operation. At the end of the section, we provide the learning algorithm for model training.

\subsection{Reasoning with Implicit Feedback}
\label{sec:logicalization}
One fundamental goal of personalized recommendation is to predict a user's future behavior given the existing behaviors. We first consider users' implicit feedbacks, i.e., we only know if a user has interacted with an item, but do not know if the user likes or dislikes the interacted item. Suppose a user $u$'s interaction history contains $r$ items $v_1,v_2,\cdots,v_r$, and we want to predict if an item $v_x$ is to be recommended for the user.
We need to define a function in first-order logic to encode the interactions into a logic space. Then the problem of recommending item $v_x$ or not reduces to the problem of deciding if the following Horn clause is True or False:
\begin{equation}
\label{eq:original}
    I(u,v_1) \wedge I(u,v_2) \wedge \cdots \wedge I(u,v_r) \rightarrow I(u,v_x)
\end{equation}

In this example, $I(u,v_i)$ is an encoding function to be learned that shows user $u$ \textit{interacted} with item $v_i$.
We can also learn this function for different meanings based on different scenarios and training data.
Intuitively, we use Horn clause to depict if the user's existing behaviors together would imply the user's preference on a new item $v_x$. In model training, each user's interaction history is represented as a logical expression. Since the number of interactions and the interacted items of different users are different, the logical expressions from all users combined represent a diverse set of training rules. Intuitively, each user contributes to part of the logical space (i.e., the user's logical expression), and they collectively estimate a reasoning model of the space to make predictions for each other, thus noted as neural collaborative reasoning. We will introduce how to learn the model in the next section.

\begin{figure}[t]
\centering
\includegraphics[scale=0.27]{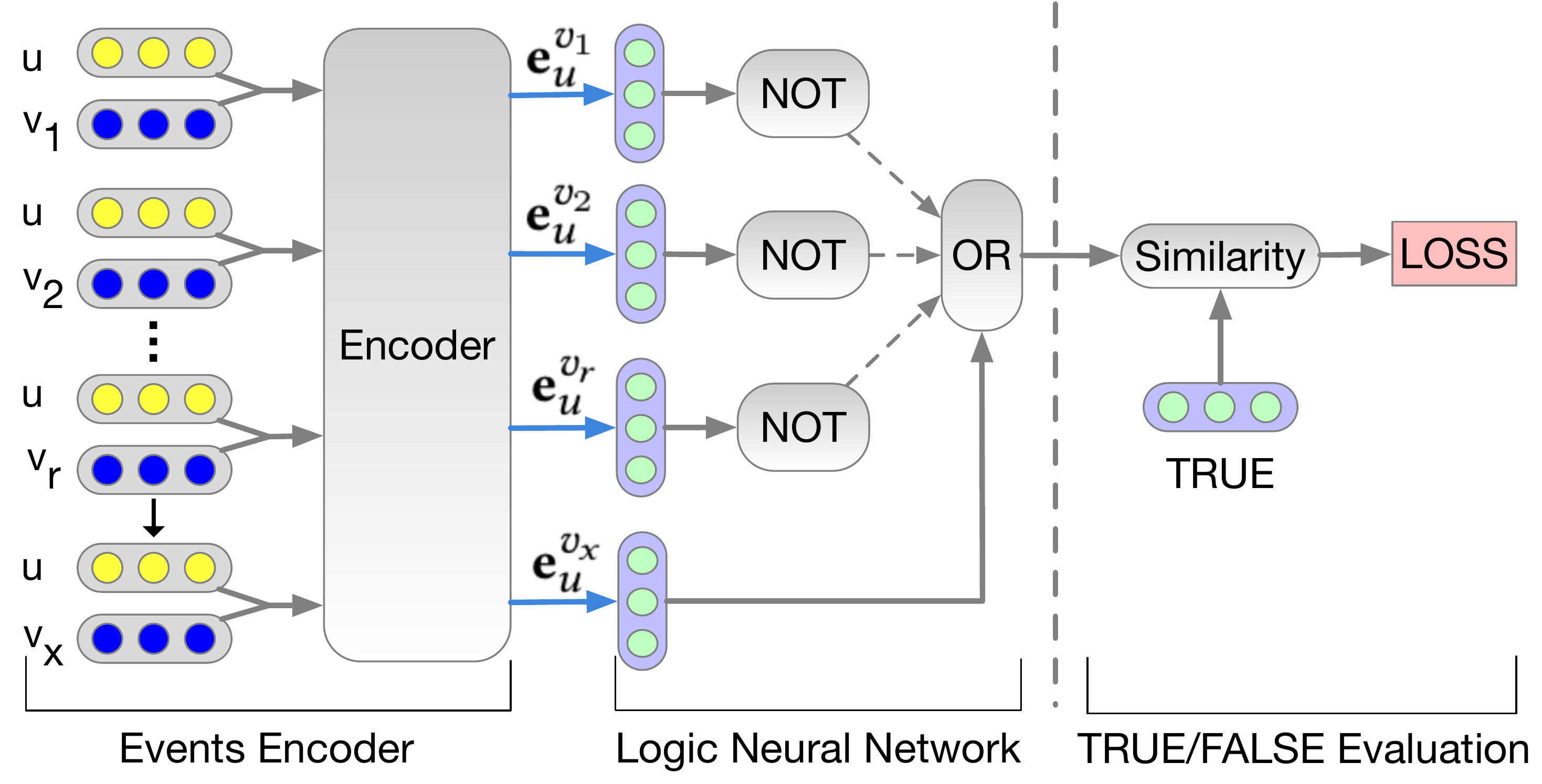}
\vspace{-5pt}
\caption{Implementation of the NCR framework. The gray boxes with yellow or blue circles represent user or item embeddings;
Blue boxes with green circles represent event embeddings in the logical space, where the encoder is a neural network that encodes user-item interactions to events; NOT and OR are neural logic modules; Dashed arrows mean that the order of the inputs are randomly shuffled in each round.}
\label{fig:ncr_model}
\vspace{-15pt}
\end{figure}

Based on the definition of material implication (Eq.\eqref{implication}), the above statement can be rewritten by only using the basic logical operations AND ($\wedge$), OR ($\vee$) and NOT ($\neg$), which is shown as follows:
\begin{equation}
\label{eq:threeoperation}
    \neg \big(I(u,v_1) \wedge I(u,v_2) \wedge \cdots \wedge I(u,v_r)\big) \vee I(u,v_x)
\end{equation}

Based on De Morgan's Law (Eq.\eqref{morgan}), this can be further rewritten into a statement using only two basic operations $\neg$ and $\vee$:
\begin{equation}
\label{eq:twooperation}
    \big(\neg I(u,v_1) \vee \neg I(u,v_2) \vee \cdots \vee \neg I(u,v_r)\big) \vee I(u,v_x)
\end{equation}

One can see that the same logical statement (e.g., Eq.\eqref{eq:original}) can be written into logically identical but literally different forms (Eq.\eqref{eq:threeoperation} and Eq.\eqref{eq:twooperation}). As a result, a natural question to ask is which form should we use to build the neural architecture. As we will show in the experiments later, it is beneficial if the neural network only needs to train two logical operation modules ($\neg$ and $\vee$) instead of all three modules ($\neg$, $\wedge$ and $\vee$). As a result, we choose Eq.\eqref{eq:twooperation} as the basic logical form to introduce the NCR framework in this section, and we will experiment with different forms in later sections.



For simplicity in notation, in the following parts of the paper, we use an \textit{event} $e$ to denote an interaction. 
Let $\mathcal{U}$ and $\mathcal{V}$ be the set of users and items.
Suppose the interaction history of user $u\in\mathcal{U}$ is $\{v_1, v_2, \cdots, v_r\}$, and then we represent these interactions $I(u,v_1), I(u, v_2), \cdots, I(u, v_r)$ as events $e_u^{v_1}, e_u^{v_2}, \cdots, e_u^{v_r}$, where $e_u^{v_i}$ means user $u$ interacted with item $v_i$. As a result, the personalized recommendation problem becomes predicting if the observed events $e_u^{v_1}, e_u^{v_2}, \cdots, e_u^{v_r}$ would imply a new event $e_u^{v_x}$ by deciding if the following statement is true:
\begin{equation}
\label{eq:implicit-horn}
    e_u^{v_1} \wedge e_u^{v_2} \wedge \cdots \wedge e_u^{v_r} \rightarrow e_u^{v_x}
\end{equation}
which, still, can be re-written using two logical operations:
\begin{equation}
\label{eq:implicit}
    (\neg e_u^{v_1} \vee \neg e_u^{v_2} \vee \cdots \vee \neg e_u^{v_r}) \vee e_u^{v_x}
\end{equation}
where $e_u^{v_x}$ is for $I(u, v_x)$. As we will illustrate later, the embedding $\textbf{e}_u^v$ for event $e_u^v$ will consider both user $u$ and item $v$, so that the model can make personalized recommendations tailored to a user. 
Note that we do not make use of the time information in the above modeling, as a result, the ordering of the observed events in the left side of Eq.\eqref{eq:implicit-horn} does not matter. Making use of the time ordering information for sequential NCR will be considered as a future work.

\subsection{Reasoning with Explicit Feedback}
Sometimes users will not only interact with items, but also will tell us if she likes or dislikes the item. Such explicit feedback signals are very informative for the recommendation task, as a result, it would be very beneficial if we can take the explicit feedback into the logical reasoning procedure.

Fortunately, it is very easy to extend the above formalization for reasoning with explicit feedback.
In particular, we change the predicate function $I(u,v)$ to $L(u,v)$, which means if the user \textit{likes} the interacted item. 
Then, we can extend the definition of event $e_u^v$ to describe the user attitude towards an interacted item. More specifically, we use $e_u^v$ to represent that user $u$ interacted with item $v$ with positive feedback, and use $\neg e_u^v$ to show that user $u$ interacted with item $v$ with negative feedback. Still take the previous example, suppose user $u$ interacted with items $v_1, v_2, \cdots, v_r$, and gave positive feedback on $v_1,v_2,\cdots$, while negative feedback on $v_r$ (could be negative feedback on more items), then if or not to recommend $v_x$ depends on the T/F value of the following statement:
\begin{equation}
    e_u^{v_1} \wedge e_u^{v_2} \wedge \cdots \wedge \neg e_u^{v_r} \rightarrow e_u^{v_x}
\end{equation}
which is equivalently written as:
\begin{equation}
\label{eq:explicitreasoning}
    (\neg e_u^{v_1} \vee \neg e_u^{v_2} \vee \cdots \vee \neg\neg e_u^{v_r}) \vee e_u^{v_x}
\end{equation}
Here, we keep the double negation on $e_u^{v_r}$ to make sure the negation modular will be adequately trained in the neural network, which will be explained with more details in the following subsection.


\begin{table*}[t]
\small
    \centering
    \caption{Logical laws and the corresponding equation that each logical module should satisfy in our neural architecture. Some of the laws are guaranteed by adding an explicit logical regularizer into the training loss function, while others are guaranteed by randomly shuffling the logic variables during model training. $Sim(\cdot, \cdot)$ represents a similarity measure function.}
    \begin{tabular}{llll}
      \toprule
      & Law & Equation & Logical Regularizer $r_i$\\
      \midrule
      \multirow{2}{*}{NOT} & Negation & $\neg \true = \false$ & $r_1=\frac{1}{|\mathcal{X}|}\sum_{\textbf{x}\in \mathcal{X}} 1 + Sim(\text{NOT}(\textbf{x}),\textbf{x})$    \\ 
      & Double Negation & $\neg (\neg x) = x$ & $r_2=\frac{1}{|\mathcal{X}|}\sum_{\textbf{x}\in \mathcal{X}} 1-Sim(\text{NOT}(\text{NOT}(\textbf{x})),\textbf{x})$    \\ 
      \midrule
      \multirow{4}{*}{AND} & Identity & $x \wedge \true = x$ & $r_3=\frac{1}{|\mathcal{X}|}\sum_{\textbf{x}\in \mathcal{X}} 1-Sim(\text{AND}(\textbf{x}, \mtrue),\textbf{x})$  \\
      & Annihilator & $x \wedge \false = \false$  & $r_4=\frac{1}{|\mathcal{X}|}\sum_{\textbf{x}\in \mathcal{X}} 1-Sim(\text{AND}(\textbf{x}, \mfalse),\mfalse)$   \\
      & Idempotence & $x \wedge x = x$  &  $r_5=\frac{1}{|\mathcal{X}|}\sum_{\textbf{x}\in \mathcal{X}} 1-Sim(\text{AND}(\textbf{x}, \textbf{x}),\textbf{x})$ \\
      & Complementation & $x \wedge \neg x = \false$  & $r_6=\frac{1}{|\mathcal{X}|}\sum_{\textbf{x}\in \mathcal{X}} 1-Sim(\text{AND}(\textbf{x}, \text{NOT}(\textbf{x})),\mfalse)$   \\
      \midrule
      \multirow{4}{*}{OR} & Identity & $x \vee \false = x$ & $r_7=\frac{1}{|\mathcal{X}|}\sum_{\textbf{x}\in \mathcal{X}} 1-Sim(\text{OR}(\textbf{x}, \mfalse),\textbf{x})$  \\
      & Annihilator & $x \vee \true = \true$  & $r_8=\frac{1}{|\mathcal{X}|}\sum_{\textbf{x}\in \mathcal{X}} 1-Sim(\text{OR}(\textbf{x}, \mtrue),\mtrue)$  \\
      & Idempotence & $x \vee x = x$  & $r_9=\frac{1}{|\mathcal{X}|}\sum_{\textbf{x}\in \mathcal{X}} 1-Sim(\text{OR}(\textbf{x}, \textbf{x}),\textbf{x})$ \\
      & Complementation & $x \vee \neg x = \true$ & $r_{10}=\frac{1}{|\mathcal{X}|}\sum_{\textbf{x}\in \mathcal{X}} 1-Sim(\text{OR}(\textbf{x}, \text{NOT}(\textbf{x})),\mtrue)$  \\
      \midrule
      \multirow{4}{*}{AND/OR}& \multirow{2}{*}{Associativity} & $x \vee (y\vee z) = (x\vee y)\vee z$ & \multirow{4}{*}{\text{Random Shuffling of Logic Variables}} \\
      & & $x \wedge (y\wedge z) = (x\wedge y)\wedge z$ & \\
      & \multirow{2}{*}{commutativity} & $x\vee y = y\vee x$ & \\
      & & $x\wedge y = y\wedge x$ & \\
      \bottomrule
    \end{tabular}
    \label{tb:regularizer}
    \vspace{-10pt}
\end{table*}

\subsection{Logical Modules}
\label{sec:logical_modules}
We have introduced how to formalize a recommendation task into a logical reasoning procedure. Now we introduce how to build the neural architecture based on the given logical expression. We use the implicit feedback case as a running example in this section, and later we will generalize to explicit feedback cases.

Suppose a user $u$ interacted with $v_1, v_2, \cdots, v_r$, and our model needs to predict if item $v_x$ would be interacted by $u$. We first use a simple two-layer neural network to encode the user and item interactions into event vectors. The following equation shows encoding a pair of user and item vectors into one event vector:
\begin{equation}
\label{eq:mlp}
    \textbf{e}_u^v = \textbf{W}_2\phi(\textbf{W}_1 \begin{bmatrix}\textbf{u} \\ \textbf{v}\end{bmatrix} + \textbf{b}_1) + \textbf{b}_2
\end{equation}
where $\textbf{u}, \textbf{v} \in \mathbb{R}^{d}$ are the user and item latent embedding vectors in $d$-dimensional space; $\textbf{W}_1, \textbf{W}_2$ and $\textbf{b}_1,\textbf{b}_2$ are the weight matrices and bias terms to be learned; $\textbf{e}_u^v\in \mathbb{R}^n$ represents the encoded event vector, and $\phi(\cdot)$ is the rectified linear unit (ReLU) activation function: $\phi(x) = \max(0, x)$.

\begin{figure}[t]
    \begin{subfigure}[b]{0.35\linewidth}
        \centering
        \resizebox{\linewidth}{!}{
    \includegraphics[scale=0.5]{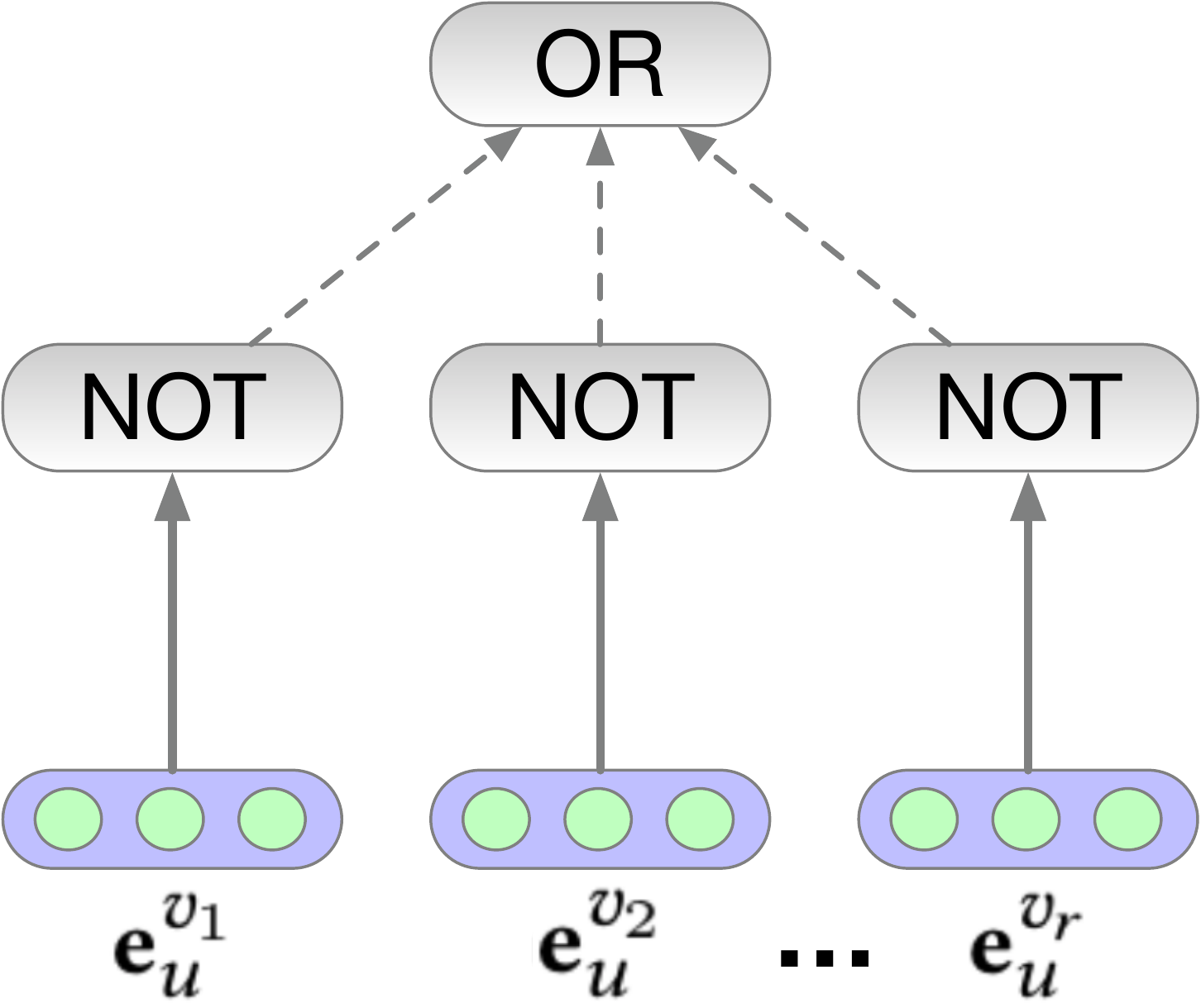}
        }
        \vspace{-10pt}
        \caption{Implicit Feedback}
        \label{fig:negative_a}
    \end{subfigure}\hspace{2em}
    \begin{subfigure}[b]{0.35\linewidth}
    \centering
        \resizebox{\linewidth}{!}{
        \includegraphics[scale=0.5]{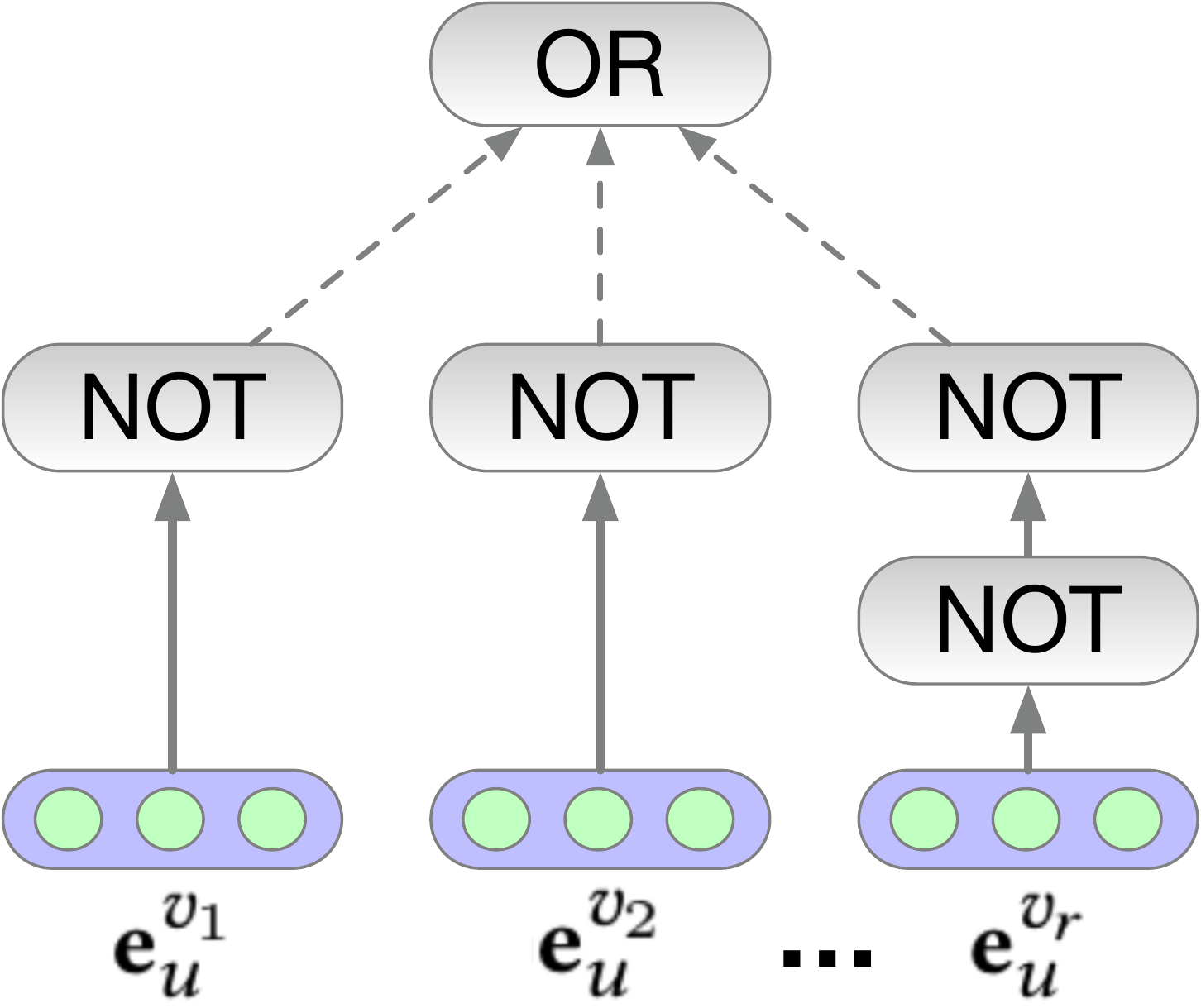}
        }
        \vspace{-10pt}
        \caption{Explicit Feedback}   
        \label{fig:negative_b}
    \end{subfigure}
\caption{Reasoning over implicit (a) and explicit (b) feedbacks. The figure only shows the Logic Operation portion of Figure \ref{fig:ncr_model}, other parts of the model are unchanged.}
\label{fig:module_not}
\vspace{-15pt}
\end{figure}

With these event vectors, the next step is to construct the neural architecture to model the logical expression in Eq.\eqref{eq:implicit}, which is now shown as vector representations based on event embeddings:
\begin{equation} \label{le_5}
    (\neg \textbf{e}_u^{v_1} \vee \neg\textbf{e}_u^{v_2} \vee \cdots \vee \neg\textbf{e}_u^{v_r}) \vee \textbf{e}_u^{v_x}
\end{equation}

Our goal is to calculate the above logical expression in a continuous representation space, and the space is characterized by two constant vectors $\mtrue$ and $\mfalse$ ($\mfalse=\neg\mtrue$). The $\mtrue$ vector is randomly initialized and kept unchanged during model training, and the $\mfalse$ vector is calculated by $\neg\mtrue$.
We expect that the final event vector of the expression would be close to $\mtrue$ if $v_x$ should be recommended, and $\mfalse$ otherwise. To achieve this goal, we represent each logical operation $\wedge,\vee,\neg$ as a neural module $\text{AND}(\cdot,\cdot), \text{OR}(\cdot,\cdot)$, and $\text{NOT}(\cdot)$, where each neural module is also a two-layer neural network (Eq.\eqref{eq:mlp}). For example, the $\text{AND}(\cdot,\cdot)$ module takes two event embeddings $\textbf{e}_1$ and $\textbf{e}_2$ as input, and outputs a new event embedding, which represents the event that both $\textbf{e}_1$ and $\textbf{e}_2$ happens.


Based on the event embeddings and logical modules, we can then assemble a neural architecture for Eq.\eqref{le_5}, as shown in Figure \ref{fig:ncr_model}.
By sending each input event embedding into the $\text{NOT}(\cdot)$ module, we can calculate the negated events $\neg\textbf{e}_u^{v_1}, \neg\textbf{e}_u^{v_2}, \cdots, \neg\textbf{e}_u^{v_r}$. After that, we combine the negated events and the candidate event embedding $\textbf{e}_u^{v_x}$ into the $\text{OR}(\cdot,\cdot)$ module, so as to generate the final event embedding of the whole logical expression. The $\text{OR}(\cdot,\cdot)$ operation can only take two event embeddings at each time, to calculate the joint embedding of more than two events using OR, we first feed in two events, e.g. $\textbf{e}_u^v$ and $\textbf{e}_u^{v^\prime}$. The output vector $\textbf{e}_u^{{v,v^\prime}}$ is treated as the representation of event $\textbf{e}_u^v \vee \textbf{e}_u^{v^\prime}$ in the logical representation space. The next event vector $\textbf{e}_u^{v^{\prime\prime}}$ and the previous output $\textbf{e}_u^{{v,v^\prime}}$ will be sent to this OR neural module again to get the embedding of three disjunction events. We conduct this recurrently until the entire OR expression is calculated. The process, which is shown in Figure \ref{fig:ncr_model}, can be represented by the following equations: 
\begin{equation}
\label{eq:ncr}
\begin{split}
    \neg\textbf{e}_u^{v_i} &= \text{NOT}(\textbf{e}_u^{v_i}), \forall i\in \{1,2,\cdots,r\}\\
    \textbf{Exp} &= \text{OR}\left(\neg\textbf{e}_u^{v_1}, \neg\textbf{e}_u^{v_2},\cdots,\neg\textbf{e}_u^{v_r},\textbf{e}_u^{v_x}\right)
\end{split}
\end{equation}

The final output $\textbf{Exp}$ is the vector representation of the logical Expression in Eq.\eqref{le_5}. To determine if the expression represents true or false, we examine if the final event embedding $\textbf{Exp}$ is close to the constant true vector (\textbf{T}) in the logical space. As stated before, the true vector is randomly initialized at the beginning and it is never updated during model training, which serves as the anchor vector for all other latent vectors in the logical space. If a vector represents true, then the vector should be close to this true vector, otherwise it should be far from the true vector. Any measure can be used to compare the $\textbf{Exp}$ and $\textbf{T}$ vectors. In this work, we use the most simple cosine similarity measure:
\begin{equation}
    Sim(\textbf{Exp},\textbf{T}) = \frac{\textbf{Exp} \cdot \textbf{T}}{\|\textbf{Exp}\| \|\textbf{T}\|}
\end{equation}

The above illustration is based on implicit feedback reasoning. To conduct reasoning based on explicit feedbacks such as Eq.\eqref{eq:explicitreasoning}, we only need to slightly modify the neural architecture in Figure \ref{fig:ncr_model}. In particular, positive events are still fed into the neural network as before, but negative events will pass through an extra $\text{NOT}(\cdot)$ module before feeding into the original architecture, as illustrated in Figure \ref{fig:module_not}. In this design, we preserve the double negation structure instead of deleting both of them to make sure the negation module can be adequately optimized in the model training procedure.

Since the number of variables (i.e., interactions) and the numbers of negative feedbacks vary for different users, the length and structure of the logical expression would be different. As a result, the neural structure are different for different users, which will be dynamically assembled according to the input expression. 

\subsection{Logical Regularization}
We have defined three logical neural modules. However, by now they are just plain neural networks. We need to guarantee that each logical module is really performing the expected logical operation in the latent space. 
To achieve this goal, similar to \cite{shi2020neural},
we add logical regularizer to the neural modules to constrain their behaviors.
The regularizers and their corresponding laws are listed in Table \ref{tb:regularizer}. 

Let $\textbf{x}$ represent an event embedding, which could be the original user-item interaction event (e.g., the $\textbf{e}_u^{v_i}$ in Eq.\eqref{eq:ncr}),
or any intermediate event during the logical neural network calculation (e.g., the output $\neg\textbf{e}_u^{v_i}$ by feeding $\textbf{e}_u^{v_i}$ to the $\text{NOT}(\cdot)$ module),
or the final event embedding \textbf{Exp} of the logical expression. Let $\mathcal{X}$ be the set of all event embeddings, and let $Sim(\cdot, \cdot)$ be a similarity function, which is cosine similarity in our implementation. As noted before, $\mtrue$ is the constant anchor vector representing true, which is randomly initialized and never updated during model learning, and $\mfalse$ is the vector representing false, which is obtained through NOT($\mtrue$).

We take the double negation rule $r_2$ as an example to explain the basic idea of logical regularizers. For the $\text{NOT}(\cdot)$ module to perform negation operation in the latent space, we require it to satisfy the double negation law: 
\begin{equation}
    \textbf{x} \equiv \text{NOT}(\text{NOT}(\textbf{x}))
\end{equation}
which means that any event embedding, if negated for twice, should return to itself. To constrain the behavior of the negation module, we design a regularizer to maximize the similarity between the output of NOT(NOT($\textbf{x}$)) and $\textbf{x}$, which is equal to minimizing: 
\begin{equation}
    1 - Sim\left(\text{NOT}(\text{NOT}(\textbf{x})), \textbf{x}\right)
\end{equation}

To make sure that the logic neural modules can not only perform the expected operation on the initial input events, but also on all of the intermediate hidden events as well as the final event, we apply the regularizer to all of the event embeddings $\mathcal{X}$ in the logical space, which gives us the final regularizer for the double negation law:
\begin{equation}
    r_2=\frac{1}{|\mathcal{X}|}\sum_{\textbf{x}\in \mathcal{X}} 1-Sim(\text{NOT}(\text{NOT}(\textbf{x})),\textbf{x})
\end{equation}
where $|\mathcal{X}|$ is the size of the entire event space. We would not introduce the details for all of the regularizers listed in Table~\ref{tb:regularizer}, since they are designed in similar ways. The only difference is the regularizer for negation law $r_1$, where we conduct one plus the similarity instead of one minus similarity, because we want to maximize the distance between NOT($\textbf{x}$) and $\textbf{x}$, such as $\mtrue$ and $\mfalse$. The final regularizer considering all laws is:
\begin{equation}
    \mathcal{L}_{logicReg} = \sum\nolimits_i r_i
\end{equation}

The associative and commutative laws cannot be easily represented as regularizers. Instead, we randomly shuffle the order of the input events every iteration during the training process to make the learned $\text{AND}$ and $\text{OR}$ modules satisfy these two laws. 


\subsection{Learning Algorithm}
In this work, we use the pair-wise learning algorithm \cite{bpr} for model training. Specifically, we conduct negative sampling on each given expression during the training process. Suppose we observed that user $u$ interacted with $r$ items $v_{i-1}, v_{i-2}, \cdots, v_{i-r}$ and then the user interacted with item $v_i$. We sample another item $v_j\in\mathcal{V}$ that the user did not interact with. Based on this, we build the structured neural network in terms of the following two expressions:
\begin{equation}
\begin{split}
    \mathcal{C}_{ui}^+ &= \neg e_u^{v_{i-1}}\vee \neg e_u^{v_{i-2}}\vee \cdots \vee \neg e_u^{v_{i-r}} \vee e_u^{v_i} \\
    \mathcal{C}_{uj}^- &= \neg e_u^{v_{i-1}}\vee \neg e_u^{v_{i-2}}\vee \cdots \vee \neg e_u^{v_{i-r}} \vee e_u^{v_j} \\
\end{split}
\end{equation}
where $\mathcal{C}_{ui}^+$ is the expression for the observed ground-truth interaction, and $\mathcal{C}_{uj}^-$ is the expression for the negative sampled interaction. Then, we have a pair of truth evaluation results:
\begin{equation}
    \begin{split}
        s_{ui}^+ &= Sim\big(\text{LNN}(\textbf{e}_u^{v_{i-1}}, \textbf{e}_u^{v_{i-2}}, \cdots, \textbf{e}_u^{v_{i-r}}, \textbf{e}_u^{v_i}), \mtrue\big) \\
        s_{uj}^- &= Sim\big(\text{LNN}(\textbf{e}_u^{v_{i-1}}, \textbf{e}_u^{v_{i-2}}, \cdots, \textbf{e}_u^{v_{i-r}}, \textbf{e}_u^{v_j}), \mtrue\big) \\
    \end{split}
\end{equation}
where LNN is the logic neural network structure as shown in Figure \ref{fig:ncr_model}. Then we calculate $s_{uij}=\alpha\cdot(s_{ui}^+ - s_{uj}^-)$ to represent the difference between these two expressions and apply an optimization algorithm to maximize this difference, where $\alpha$ is an amplification factor to amplify the difference ($\alpha=10$ in our experiments). In implementation, we sample $n$ negative items for each user-item pair. We use $\mathcal{V}_u^+$ to represent the observed example set for user $u$, and $\mathcal{V}_u^-$ to represent the sampled negative example set for user $u$, where $v_i\in \mathcal{V}_u^+$ and $v_j\in \mathcal{V}_u^-$. The loss function can be written as:
\begin{equation}
\label{eq:l_ncr}
    \mathcal{L}_{ncr} = -\sum_{u\in\mathcal{U}}\sum_{v_i\in \mathcal{V}_u^+}\sum_{v_j\in \mathcal{V}_u^-} \ln{\sigma(s_{uij}}) + \lambda_\Theta\|\Theta\|_2^2
\end{equation}
where $\Theta$ represents all of the parameters of the model, including the user and item embeddings, the parameters of the event encoder network, and the parameters of the neural modules; $\lambda_\Theta$ is the $\ell_2$-norm regularization coefficient; $\sigma(\cdot)$ is the logistic sigmoid function: $\sigma(x) = \frac{1}{1+e^{-x}}$. Maximizing $s_{uij}$ is equivalent to minimizing $\mathcal{L}_{ncr}$. The pseudo-code for calculating the logic neural network loss is given in Appendix \ref{ap:algorithm}. 

Now we can integrate the logic regularizer together with our pairwise learning loss to get the final loss function:
\begin{equation}\label{eq:model_los}
    \mathcal{L} = \mathcal{L}_{ncr} + \lambda_r \mathcal{L}_{logicReg}
\end{equation}
where $\lambda_r$ is the coefficient for the logic regularizers. We apply the same coefficient to all of the logic regularizers since they are equally important to regularize the logical behavior of the model. We apply back propagation \cite{rumelhart1986learning} to optimize the parameters.

\section{Experiments}\label{sec:experiment}
As the key motivation of the work is to develop a novel neural collaborative reasoning framework to harness the power of learning and reasoning for personalized recommendation, we aim to answer the following research questions in the experiments.
\begin{itemize}
    \item \textbf{RQ1}: What is the performance of the NCR framework in terms of personalized recommendation tasks? Does it outperform state-of-the-art models? (Section \ref{sec:results_1})
    \item \textbf{RQ2}: If and how does the logic regularizer help to improve the performance? (Section \ref{sec:results_2})
    \item \textbf{RQ3}: Does the logical prior over the neural network structure help to improve the performance? (Section \ref{sec:results_3})
    \item \textbf{RQ4}: Can we model the recommendation problem with pure Boolean logic? (Section \ref{sec:results_4})
\end{itemize}

\subsection{Experiment Dataset}
We experiment with three publicly available datasets. The statistics of the datasets are summarized in Table \ref{tb:dataset}. The size of these three datasets ranges from 10K up to million level, and they cover movies as well as e-commerce recommendation scenarios. 

\vspace{-5pt}
\begin{table}[H]
\caption{Statistics of the datasets in our experiments.}\label{tb:dataset}
\begin{tabular}{|c|c|c|c|c|}
\hline
Dataset& \#Users &\#Items &\#Interaction & Density\\
\hline
ML100k & 943 &1,682 &100,000& 6.30\%\\ 
\hline
Movies \& TV &123,961 &50,053 &1,697,533 &0.027\%\\
\hline
Electronics &192,404 &63,002 &1,689,188 & 0.014\%\\
\hline
\end{tabular}
\vspace{-10pt}
\end{table}

\textbf{ML100k}
~\cite{harper2016movielens}. This is a frequently used dataset maintained by Grouplens. It includes 100,000 movie ratings ranging from 1 to 5 from 943 users to 1,682 movies.


\textbf{Amazon 5-core}
~\cite{mcauley2015image}. This is the Amazon e-commerce dataset, which includes user, item and rating information spanning from May 1996 to July 2014. Compared with ML100k, this is a relatively sparse dataset. It covers 24 different categories, and we take \textbf{Movies and TV} and \textbf{Electronics}, which are two million-scale datasets.

Our NCR framework can be implemented in two ways: reasoning with implicit feedback or with explicit feedback. Following common practice, when reasoning with implicit feedback, we only consider the user interaction information and ignore the ratings, while for reasoning with explicit feedback, we consider 1-3 ratings as negative feedback and 4-5 ratings as positive feedback. 

Since our baseline models include some sequential recommendation models, according to the suggestions of \cite{zhao2020revisiting}, we use leave-one-out setting under temporal ordering, i.e., the last interaction of each user is put into the test set, the second-to-last interaction of each user is put into the validation set, and other interactions of each user constitute the training set.
Details of the dataset pre-processing procedure are provided in the Appendix \ref{ap:datasets}.


\subsection{Baselines}
According to the suggestions in \cite{dacrema2019we,rendle2020neural}, we consider both shallow and deep models as baselines.
We compare with two representative shallow models (BPR-MF and SVD++), two deep models (DMF and NeuMF), two session-based models (GRU4Rec and STAMP), as well as one state-of-the-art reasoning-based model (NLR).
\begin{itemize}
    \item \textbf{BPR-MF} \cite{bpr}: The Bayesian Personalized Ranking model, which is a pair-wise ranking model for recommendation. We use Biased Matrix Factorization (Bias-MF) \cite{koren2009matrix} as the prediction function under the BPR framework, which considers user, item and global bias terms for matrix factorization. We denote the final model as BPR-MF.
    \item \textbf{SVD++} \cite{koren2008factorization}: Also a matrix factorization based method, which extends Singular Value Decomposition (SVD) by considering user history interactions when modeling the users.
    \item \textbf{DMF} \cite{xue2017deep}: Deep Matrix Factorization is a deep model for recommendation, which uses multiple non-linear layers to process the raw user-item interaction matrix. 
    \item \textbf{NeuMF} \cite{he2017neural}: A neural network-based collaborative filtering algorithm, which employs a non-linear prediction network for user and item matching. 
    \item \textbf{GRU4Rec} \cite{hidasi2016session}: A sequential/session-based recommendation model, which uses Recurrent Neural Network (RNN) to capture the sequential dependencies in users' historical interactions for prediction and recommendation.
    \item \textbf{STAMP} \cite{liu2018stamp}: A sequential/session-based recommendation model based on the attention mechanism, which captures a user's long-term and short-term preferences.
    \item \textbf{NLR} \cite{shi2020neural}: Neural Logic Reasoning, which proposes a Logic-Integrated Neural Network (LINN) to take logical constraints and neural modeling for reasoning and prediction.
\end{itemize}

In the experiments, we test three versions of our model:
\begin{itemize}
    \item \textbf{NCR-I}: Neural Collaborative Reasoning with Implicit feedback, which only uses the interaction information for model learning.
    \item \textbf{NCR-E}: Neural Collaborative Reasoning with Explicit feedback, which adopts the explicit feedback for model learning.
    \item \textbf{NCR-E w/o LR}: Neural Collaborative Reasoning with Explicit feedback but without Logical Regularization, i.e., we remove the logical regularizers from NCR-E by setting $\lambda_r=0$ in Eq.\eqref{eq:model_los}.
\end{itemize}

We train the pair-wise ranking methods based on 1:1 negative sampling, i.e., for each interacted item $v_i\in \mathcal{V}_u^+$, we sample one negative item $v_j\in \mathcal{V}_u^-$ that the user did not interact with. Source code of our model and the baselines are available on GitHub.\footnote{\url{https://github.com/rutgerswiselab/NCR}}

\subsection{Evaluation Metrics}
To evaluate the top-$K$ recommendation performance, we use standard metrics such as Normalized Discounted Cumulative Gain at rank $K$ (NDCG@$K$) and Hit Ratio at rank $K$ (HR@$K$). In our experiments, the result of all metrics are averaged over all users.

According to the suggestions of \cite{zhao2020revisiting}, we use \textit{real-plus-N} \cite{bellogin2011precision,said2014comparative} to calculate the measures. More specifically, for each user-item pair in the validation and test set, we randomly sample 100 irrelevant items, and we rank these 101 items for ranking evaluation.

\subsection{Experimental Settings}
We use the same train, validation and test datasets for our model and baseline methods in experiments. For fair comparison, for all models including our model and baselines, we tune each model's parameter to its own best performance on the validation set based on NDCG@5. Eventually, all other models except for DMF are set with an embedding size of 64, while for DMF, the embedding size is 128. More details of the parameter settings are shown in Appendix \ref{ap:baseline}. For our NCR model, the number of layers for all neural modules are set to 2. We apply ReLU non-linear activation function between layers. The learning rate is 0.001, and the weight of $\ell_2$-regularization coefficient ($\lambda_\Theta$ in Eq.\eqref{eq:l_ncr}) is 0.0001 for ML100k dataset and 0.00001 for Amazon datasets. The default logical regularization coefficient ($\lambda_r$ in Eq.\eqref{eq:model_los}) that we use to report the results is 0.1, and we tune the parameter to see its effect in Section \ref{sec:results_2}. We optimize the models using mini-batch Adam~\cite{kingma2014adam} with a batch size of 128. More implementation details of the models, as well as the hardware and software settings are provided in Appendix \ref{ap:baseline}.


\pgfplotsset{
        compat=1.3,
        my axis style/.style={
            every axis plot post/.style={/pgf/number format/fixed,
            /pgf/number format/precision=4},
            ybar=5pt,
            bar width=8pt,
            height=5.5cm,
            x=1.2cm,
            axis on top,
            enlarge x limits=0.2,
            symbolic x coords={16,32,64,128},
            visualization depends on=rawy\as\rawy, 
            nodes near coords={%
                \pgfmathprintnumber[precision=4]{\rawy}
            },
            every node near coord/.append style={rotate=90, anchor=west},
            label style={font=\normalsize},
            xtick distance=1,
        },
    }

\begin{table*}[t]
\setlength{\tabcolsep}{2pt}
\caption{Results of recommendation performance on three datasets with metrics NDCG (N) and Hit Ratio (HR). We use underline (\underline{number}) to show the best result among the matching-based baselines (i.e., the first six baselines), and we use underwave (\uwave{number}) to show the best result among all baselines including the reasoning-based NLR method. We use bold font to mark the best result of the whole column. We use one star (*) to indicate that the performance is significantly better than the best matching-based baselines, and use two stars (**) to indicate that the performance is significantly better than all baselines including the NLR baseline. The significance is at 0.05 level based on paired $t$-test. Improvement$^1$ shows our model improvement over the best matching-based result (i.e., over \underline{number}), while improvement$^2$ shows our model improvement over NLR.}\label{tb:results}
\centering\begin{tabular}{lcccccccccccc}
\toprule
\multirow{2}{*}{} &\multicolumn{4}{c}{\textbf{ML100k}} & \multicolumn{4}{c}{\textbf{Movies and TV}} & \multicolumn{4}{c}{\textbf{Electronics}} \\ 
\cmidrule(lr){2-5}
\cmidrule(lr){6-9}
\cmidrule(lr){10-13}
&N@5 &N@10 &HR@5 &HR@10 &N@5 &N@10 &HR@5 & HR@10 &N@5 &N@10 &HR@5 &HR@10 \\
\midrule
 BPR-MF &0.3024 &0.3659 &0.4501 &0.6486 &0.3962 &0.4392 &0.5346 &0.6676 &0.3092 &0.3472 &0.4179 &0.5354\\
 SVD++ & 0.3087 & 0.3685 &0.4586 &0.6433 &0.3918 &0.4335 &0.5224 &0.6512 &0.2775 &0.3172 &0.3848 &0.5077\\
 \midrule
 DMF &0.3023 &0.3661 &0.4480 & 0.6450 &0.4006 &0.4455 &\underline{0.5455} & \uwave{\underline{0.6843}} & 0.2775 &0.3143 & 0.3783 & 0.4922 \\
 NeuMF &0.3002 &0.3592 &0.4490 &0.6316 & 0.3791 & 0.4211 &0.5134 & 0.6429 & 0.3026 & 0.3358 &0.4031 & 0.5123\\ 
 \midrule
 GRU4Rec &\underline{0.3564} &\underline{0.4122} &0.5134 &\uwave{\underline{0.6856}} &\underline{0.4038} &\underline{0.4459} &0.5287 &0.6688 &\underline{0.3154}& \underline{0.3551} & \underline{0.4284} & \underline{0.5511}\\ 
 STAMP &0.3560 &0.4070 &\uwave{\underline{0.5159}} &0.6730 &0.3935 &0.4366 &0.5246 &0.6577 &0.3095&0.3489 &0.4196 &0.5430 \\ 
 \midrule
 NLR &\uwave{0.3602} &\uwave{0.4151} &0.5102 &0.6795 & \uwave{0.4191} & \uwave{0.4591} & \uwave{0.5506} & 0.6739 & \uwave{0.3475} & \uwave{0.3852} & \uwave{0.4623} & \uwave{0.5788} \\
 \midrule
\textbf{NCR-I} &0.3697 &0.4219 &0.5265 &0.6890 & 0.4152 & 0.4550 & 0.5479 & 0.6709 &0.3226 &0.3604 &0.4331 &0.5500 \\ 
\textbf{NCR-E} w/o \textbf{LR} &0.3671 &0.4219 &0.5180 &0.6890 &0.4126 &0.4535 &0.5444 &0.6705 &0.3272 &0.3649 &0.4377 &0.5544 \\
\textbf{NCR-E} &\textbf{0.3760**} &\textbf{0.4240**} &\textbf{0.5456**} &\textbf{0.6943**} & \textbf{0.4255**} & \textbf{0.4670**} &\textbf{0.5611**} & \textbf{0.6891} &\textbf{0.3499*} &\textbf{0.3878*} &\textbf{0.4639*} &\textbf{0.5812*} \\ 
 \midrule
 \midrule
Improvment$^1$ & 5.50\% & 2.86\% & 5.76\% & 1.27\% & 5.37\% & 4.73\% & 2.86\% & 0.70\% & 10.94\% & 9.21\% & 8.29\% & 5.46\%\\

Improvment$^2$ & 4.39\% & 2.14\% & 6.71\% & 2.66\% & 1.53\% & 1.72\% & 1.91\% & 2.26\% & 0.69\% & 0.67\% & 0.35\% & 0.41\%\\
 \bottomrule
\end{tabular}
\vspace{-5pt}
\end{table*}

\begin{figure*}
    \begin{subfigure}[b]{0.29\textwidth}
        \centering
        \resizebox{0.8\linewidth}{!}{
\begin{tikzpicture}
\pgfplotsset{
    scale only axis,
    xmin=0, xmax=1,
     label style={font=\normalsize}
}

\begin{axis}[
  axis y line*=left,
  ymin=0.39, ymax=0.43,
  xlabel={Logical Regularization Coefficient},
  ylabel={NDCG@10},
  xticklabels={0,0,$10^{-4}$,$10^{-3}$,$10^{-2}$,$10^{-1}$,1},
  legend pos=south west,
  ymajorgrids=true
]
\addplot[mark=square,red]
  coordinates{
    (0,0.4219)(0.2,0.4199)(0.4,0.4205)(0.6,0.4030)(0.8,0.4240)(1.0,0.4003)
}; \label{plot_one}
\end{axis}

\begin{axis}[
  axis y line*=right,
  axis x line=none,
  ymin=0.66, ymax=0.74,
  ylabel={HR@10},
]
\addplot[mark=*,blue]
  coordinates{
    (0,0.6890)
    (0.2,0.6858)
    (0.4,0.6879)
    (0.6,0.6709)
    (0.8,0.6943)
    (1.0,0.6667)
}; 
\end{axis}
\end{tikzpicture}
        }
        \caption{ML100k}
        \label{fig:subfig8}
    \end{subfigure}
    \hskip -8.5ex
    \begin{subfigure}[b]{0.29\textwidth}
    \centering
        \resizebox{0.8\linewidth}{!}{
            \begin{tikzpicture}
\pgfplotsset{
    scale only axis,
    xmin=0, xmax=1,
    label style={font=\normalsize}
}

\begin{axis}[
  axis y line*=left,
  ymin=0.405, ymax=0.475,
  xlabel={Logical Regularization Coefficient},
  ylabel={NDCG@10},
  xticklabels={0,0,$10^{-4}$,$10^{-3}$,$10^{-2}$,$10^{-1}$,1},
  legend pos=south west,
  ymajorgrids=true,
  legend style={font=\fontsize{7}{5}\selectfont}
]
\addplot[mark=square,red]
  coordinates{
  (0,0.4535)(0.2,0.4458)(0.4,0.4520)(0.6,0.4438)(0.8,0.4670)(1,0.4202)

}; \label{plot_one}
\end{axis}

\begin{axis}[
  axis y line*=right,
  axis x line=none,
  ymin=0.630, ymax=0.700,
  ylabel={HR@10},
  xticklabels={0,0,$10^{-4}$,$10^{-3}$,$10^{-2}$,$10^{-1}$,1},
]
\addplot[mark=*,blue]
  coordinates{
    (0,0.6705)
    (0.2,0.6650)
    (0.4,0.6688)
    (0.6,0.6600)
    (0.8,0.67891)
    (1,0.6392)
}; 
\end{axis}
\end{tikzpicture}
        }
        \caption{Movies and TV}   
        \label{fig:subfig9}
    \end{subfigure}
    \hskip -8.5ex
    \begin{subfigure}[b]{0.29\textwidth}
        \centering
        \resizebox{0.8\linewidth}{!}{
            \begin{tikzpicture}
\pgfplotsset{
    scale only axis,
     label style={font=\normalsize},
    xmin=0, xmax=1
}

\begin{axis}[
  axis y line*=left,
  ymin=0.325, ymax=0.395,
  xlabel={Logical Regularization Coefficient},
  ylabel={NDCG@10},
  xticklabels={0,0,$10^{-4}$,$10^{-3}$,$10^{-2}$,$10^{-1}$,1},
  ymajorgrids=true,
]
\addplot[mark=square,red]
  coordinates{
    (0,0.3649)(0.2,0.3561)(0.4,0.3628)(0.6,0.3496)(0.8,0.3878)(1,0.3545)
}; \label{plot_one}
\end{axis}

\begin{axis}[
  axis y line*=right,
  axis x line=none,
  ymin=0.525, ymax=0.595,
  xticklabels={0,0,$10^{-4}$,$10^{-3}$,$10^{-2}$,$10^{-1}$,1},
  ylabel={HR@10},
]
\addplot[mark=*,blue]
  coordinates{
    (0,0.5544)
    (0.2,0.5439)
    (0.4,0.5499)
    (0.6,0.5360)
    (0.8,0.5812)
    (1,0.5446)
    
}; 
\end{axis}
\end{tikzpicture}
        }
        \caption{Electronics}
        \label{fig:subfig10}
    \end{subfigure}
    \hskip -11ex
\begin{subfigure}[b]{0.29\textwidth}
        \centering
        \resizebox{0.675\linewidth}{!}{
            \begin{tikzpicture}
\pgfplotsset{
    scale only axis,
     label style={font=\normalsize},
    xmin=0, xmax=0.8
}

\begin{axis}[
  ymin=0.2350, ymax=0.4450,
  xlabel={Event Embedding Loss Coefficient},
  ylabel={NDCG@10},
  xticklabels={0,0,$10^{-3}$,$10^{-2}$,$10^{-1}$,1},
  ymajorgrids=true,
]
\addplot[mark=square,red]
 coordinates {
    (0,0.4320)
    (0.2,0.3859)
    (0.4,0.3192)
    (0.6,0.2520)
    (0.8,0.2756)
}; \label{plot_boolean_logic}
\end{axis}
\end{tikzpicture}
        }
        \caption{$\lambda_e$ on ML100k}
        \label{fig:subfig10}
    \end{subfigure}

\caption{(a)-(c): NDCG@10 (red squared line) and HR@10 (blue circled line) on three datasets according to the increase of the logical regularization coefficient $\lambda_r$. (d): NDCG@10 when increasing the event embedding loss coefficient $\lambda_e$ on ML100k.}
\vspace{-10pt}
\label{fig:logic_regularizer_weight}
\end{figure*}
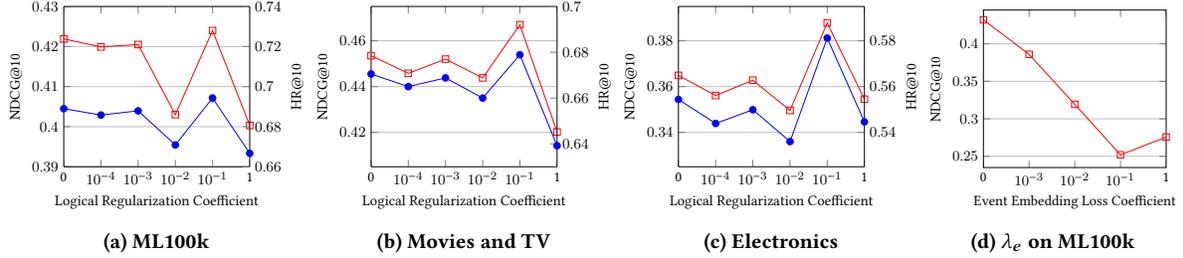

\subsection{Performance of the NCR Framework (RQ1)}\label{sec:results_1}
To evaluate the performance of the NCR framework, we report the results, including NDCG and Hit-Ratio (HR), on all of the datasets in Table~\ref{tb:results}, where \textbf{NCR-I} represents our model with implicit feedback, and \textbf{NCR-E} represents our model with explicit feedback. Besides, we use underline to highlight the best result among the matching-based baselines (i.e., the first six baselines) in each column, and we use underwave to highlight the best result among all baselines, including the reasoning-based NLR baseline. Bold number shows the best result of the whole column.

We first compare with the matching-based models. We see that among the first six baselines, GRU4Rec and STAMP achieve the best performance in most cases. Since the two models use implicit feedback for model training, for fairness in comparison, we use our implicit model NCR-I to compare with the two baselines. The results show that NCR-I is better than the two baselines in most cases.
Notice that our model shuffles the input variables in every epoch, which means that we actually did not use the item ordering information. However, our neural collaborative reasoning approach can still outperform the two session-based models on most of the measures. When using explicit feedback for model learning, we see that our NCR-E model achieves even better performance. The Improvement$^1$ row in Table \ref{tb:results} shows the percentage improvement of NCR-E against the best matching-based model.

We then compare with the reasoning-based baseline, i.e, the NLR model. Since NLR uses explicit feedback for model learning, we use our explicit model NCR-E to compare with NLR. We see that NCR-E is better than NLR on all metrics. The underlying reason may be two fold. First is that NLR is a non-personalized model, which does not learn personalized user embeddings, but only relies on the logical relationship among items for recommendation. However, our NCR model is personalized by considering both user and item in the event embedding. Another reason is that NCR adopts Horn clauses for reasoning, which is a more natural and straightforward logical language for prediction tasks such as recommendation. The Improvement$^2$ row in Table \ref{tb:results} shows the percentage improvement of NCR-E against the NLR model.

\subsection{The Effect of Logical Regularization (RQ2)}\label{sec:results_2}
In this section, we answer the question that if the logical regularizers would help to improve the performance. We tune the logical regularizer coefficient $\lambda_r$ in Eq.\eqref{eq:model_los} from $10^{-5}$ to 1. The corresponding NDCG@10 and HR@10 are shown in Figure~\ref{fig:logic_regularizer_weight}(a)-(c). We can see that the best performance would be reached by assigning the logical regularization coefficient to 0.1. 
This result shows that it is useful to apply logical constraints to the neural networks to improve the recommendation performance. However, the constraints need to be carefully adjusted. If the constraint is too weak or too strong, the performance of the model would be negatively influenced.

In Table~\ref{tb:results}, $\textbf{NCR-E w/o LR}$ shows the performance of our explicit feedback model without using logical regularizers (i.e., setting $\lambda_r=0$). By comparing $\textbf{NCR-E w/o LR}$ and $\textbf{NCR-E}$, we can see that the recommendation performance improves by using logical regularizers. We also conduct paired $t$-test between the two models, and the improvements are significant at 0.05 level except for NDCG@10 on the ML100k dataset. This result shows that logical regularizers do help to improve the recommendation performance in our framework.

\begin{table*}[t]
\caption{Ranking performance under different logical structures. ``*'' indicates significance at 0.05 level under paired t-test.}\label{tb:logic_structure}
\centering\begin{tabular}{lcccccccccccc}
\toprule
\multirow{2}{*}{} &\multicolumn{4}{c}{\textbf{ML100k}} & \multicolumn{4}{c}{\textbf{Movies and TV}} & \multicolumn{4}{c}{\textbf{Electronics}} \\ 
\cmidrule(lr){2-5}
\cmidrule(lr){6-9}
\cmidrule(lr){10-13}
&N@5 &N@10 &HR@5 &HR@10 &N@5 &N@10 &HR@5 & HR@10 &N@5 &N@10 &HR@5 &HR@10 \\
\midrule
GRU4Rec &0.3564 &0.4122 &0.5134 &0.6856 &0.4038 &0.4459 &0.5287 &0.6688 &0.3154 &0.3551 &0.4284 &0.5511\\ 
NLR &0.3529 &0.4066 &0.5113 &0.6763 & 0.4191 & 0.4591 & 0.5506 & 0.6739 & 0.3475 & 0.3852 & 0.4623 & 0.5788 \\
\hline
$^1$EqModel &0.3664 &0.4224 &0.5318 &\textbf{0.7070} &0.4105 &0.4521 &0.5429 &0.6686 &0.3249 &0.3626 &0.4355 &0.5518\\
$^2$CMPModel &0.3551 &0.4144 &0.5106 &0.6932 &0.4100 &0.4506 &0.5417 &0.6670 &0.3165 &0.3541 &0.4252 &0.5416\\
$^3\textbf{NCR-E}$ &\textbf{0.3760} &\textbf{0.4240} &\textbf{0.5456} &0.6943 & \textbf{0.4255} & \textbf{0.4670} &\textbf{0.5611} & \textbf{0.6891} &\textbf{0.3499} &\textbf{0.3878} &\textbf{0.4639} &\textbf{0.5812} \\ 
\hline
\hline
$p$-value$^{1,3}$ &0.0825 &0.0606 &0.1073 &0.0547 &0.0156* &0.0230* &0.0212* &0.0197* &0.0015* &0.0021* &0.0010* &0.0009*  \\
$p$-value$^{2,3}$ &0.0099* &0.0250* &0.0258* &0.4668 &0.0108* &0.0103* &0.0057* &0.0048* &0.0022* &0.0019* &0.0023* &0.0018* \\
 \bottomrule
\end{tabular}
\vspace{-5pt}
\end{table*}

\subsection{\mbox{The Effect of Logic Prior over Structure (RQ3)}}\label{sec:results_3}


Our logical neural network structure is characterized by two important features: modularity and logical regularization. Modularity means that we dynamically assemble the neural structure according to the logical expression. Each network module is responsible for a specific operation, and the entire network structure varies in terms of the logical experessions. As a result, different user and item interaction histories would result in different network structures during both training and testing, and this is a big difference between our framework and many traditional deep learning models whose network structures are static. 

As we mentioned in Section \ref{sec:logicalization}, the same logical statement (e.g., Eq.\eqref{eq:original}) can be written into logically identical but literally different expressions (Eq.\eqref{eq:threeoperation} and \eqref{eq:twooperation}), and different expressions will result in different network structures in our model. In the previous modeling and experiments, we used the two operation ($\neg,\vee$) expression to build the network structure (Figure \ref{fig:ncr_model}). However, it would be interesting to see what happens if we use other logically identical but literally different expressions to build the network structure.

\begin{figure}[t]
    \begin{subfigure}[b]{0.40\linewidth}
        \centering
        \resizebox{\linewidth}{!}{
    \includegraphics[scale=0.2]{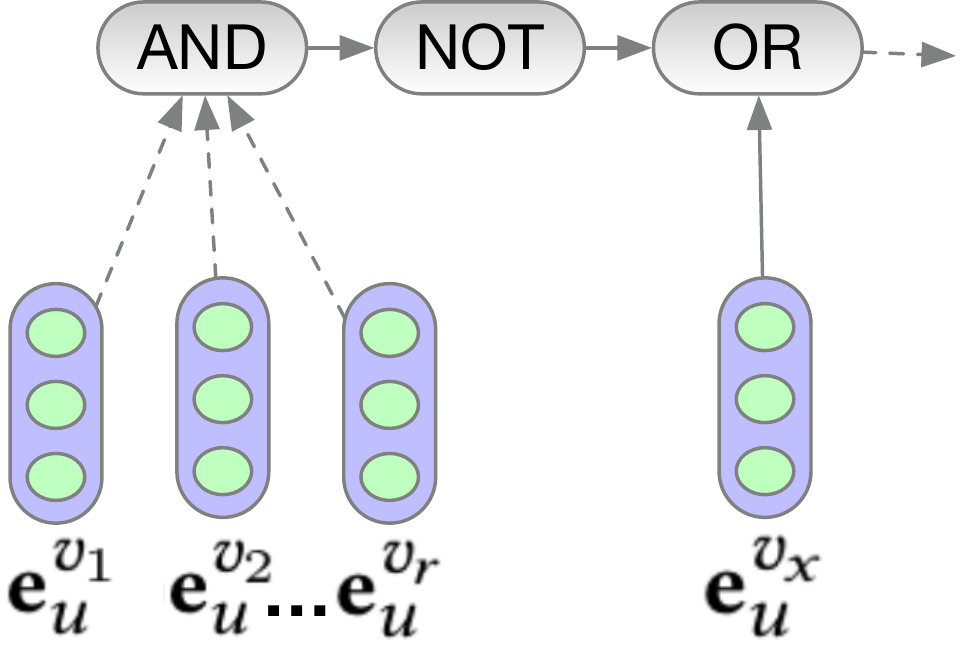}
        }
        \caption{EqModel}
        \label{fig:comparison_a}
    \end{subfigure}\hspace{2em}
    \begin{subfigure}[b]{0.32\linewidth}
    \centering
        \resizebox{\linewidth}{!}{
        \includegraphics[scale=0.3]{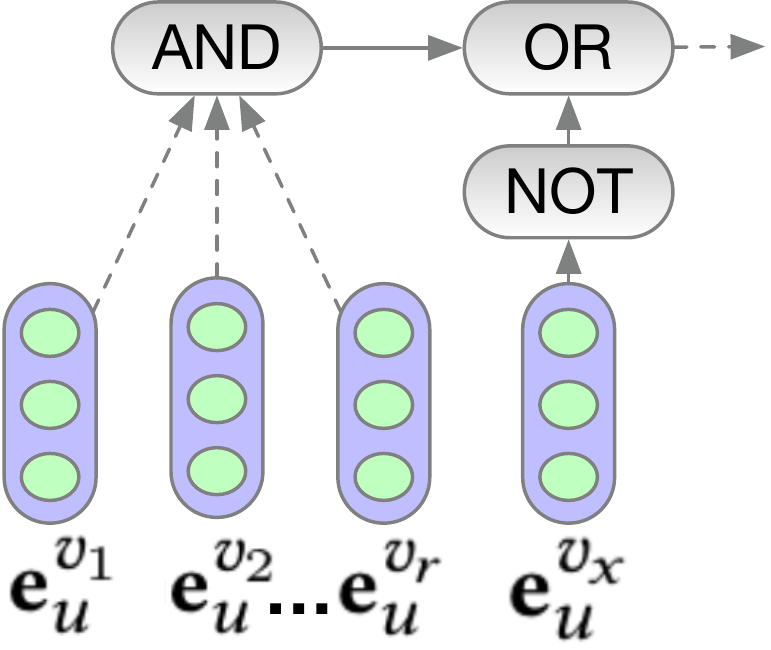}
        }
        \caption{CMPModel}
        \label{fig:comparison_b}
    \end{subfigure}
\caption{Comparison between the network structure that (a) follows the logical prior and (b) violates the logical prior.}
\label{fig:comparison}
\vspace{-10pt}
\end{figure}

To answer the question, we explore two alternative network structures. One is a logically equivalent model (\textbf{EqModel}), i.e., we still use the logical expression
$\textbf{e}_u^{v_1} \wedge \textbf{e}_u^{v_2} \wedge \cdots \wedge \textbf{e}_u^{v_r} \rightarrow \textbf{e}_u^{v_x}$
to model the task. However, it is represented as $\neg(\textbf{e}_u^{v_1} \wedge~\textbf{e}_u^{v_2}~\wedge~\cdots \wedge~\textbf{e}_u^{v_r})~\vee~\textbf{e}_u^{v_x}$, instead of 
$(\neg \textbf{e}_u^{v_1} \vee \neg\textbf{e}_u^{v_2} \vee \cdots \vee \neg\textbf{e}_u^{v_r}) \vee \textbf{e}_u^{v_x}$
that we used before (Eq.\eqref{le_5}). 
Figure~\ref{fig:comparison}(a) shows the network structure of the logically equivalent model. One can see that although this model is logically equivalent to NCR, the neural structures are different. Besides, the original network only needs to train two modules ($\neg,\vee$), while the new network needs to train all three modules ($\neg,\wedge,\vee$).

Another model is a logically nonequivalent model, noted as a comparative model (\textbf{CMPModel}). We apply the logical expression $ \textbf{e}_u^{v_x}\rightarrow\textbf{e}_u^{v_1}~\wedge~\textbf{e}_u^{v_2} \wedge \cdots \wedge \textbf{e}_u^{v_r}$, which is equivalent to $\neg\textbf{e}_u^{v_x} \vee~ (\textbf{e}_u^{v_1} \wedge \textbf{e}_u^{v_2}~\wedge \cdots \wedge~\textbf{e}_u^{v_r})$, to build the neural structure. Figure~\ref{fig:comparison}(b) shows the network structure of the CMPModel. One can see that the model attempts to use future events to predict the previous events, which violates our logical intuition about the recommendation task. 

In Table~\ref{tb:logic_structure}, based on 5-round random experiments, we provide the $p$-value under paired $t$-test between the EqModel and the original NCR-E model ($p$-value$^{1,3}$ in the table), as well as between the CMPModel and NCR-E ($p$-value$^{2,3}$ in the table). 
For easy reference, we copy the results of GRU4Rec, which is the best matching-based baseline model, as well as the results of NLR, which is the reasoning-based model from Table \ref{tb:results} to Table \ref{tb:logic_structure}.

We have two key observations from the results in Table \ref{tb:logic_structure}. First, we see that both NCR-E and the EqModel consistently outperform the GRU4Rec baseline, while the CMPModel is generally not better than the baseline. Besides, the CMPModel is significantly worse than the original NCR-E model (shown by $p$-value$^{2,3}$ in the table). This observation shows that a correct and reasonable logical structure is important to the performance of the model.

Another observation comes by comparing NCR-E and EqModel. By looking at the $p$-value between the two models (i.e., $p$-value$^{1,3}$), we see that the two models are comparable on ML100k dataset (i.e., NCR-E is not significantly better than EqModel), while NCR-E is indeed significantly better than EqModel on the two Amazon datasets. The underlying reason may arise from two factors---the complexity of model, and the sufficiency of data. As shown in Table \ref{tb:dataset}, the MovieLens dataset is 2 magnitudes denser than the Amazon datasets. Since NCR-E and EqModel are logically equivalent, they achieve comparable performance when the training data is sufficient. However, NCR-E only needs to train two neural models ($\neg,\vee$), while EqModel has to train three modules ($\neg,\wedge,\vee$), thus EqModel has a higher model complexity than NCR-E. As a result, NCR-E achieves better performance when the training data is sparse.


From this experiment, we can learn that it is important to use a reasonable logical prior to construct the model for a specific task. In addition, when there are multiple logically equivalent structures, we tend to use a simpler network structure (i.e., fewer modules) instead of a complex one.



\subsection{Boolean Logic Modeling (RQ4)}\label{sec:results_4}

In our model, we did not apply constraints to the event embeddings, as a result, they can be learned as flexible vectors in the logical space. In this experiment, to explore if the recommendation task can be modeled based on Boolean logic, we apply a constraint that any event embedding can only be either the $\mtrue$ vector or the $\mfalse$ vector. To do so, we assume that the encoder network is a prediction network, which predicts if a user would give positive feedback to an item.

We first concatenate the user $u$ and item $v$'s embeddings, and feed it into the encoder network. Before further feeding the encoded event embedding $\textbf{e}_u^v$ into the logic neural network, we calculate the mean square error (MSE) between this event embedding and the $\mtrue$ or $\mfalse$ vector, where the MSE between two vectors $\textbf{x}$ and $\textbf{y}$ is defined as $\text{MSE}(\textbf{x},\textbf{y})=\frac{1}{n}\|\textbf{x}-\textbf{y}\|_2^2$ ($n$ is the dimension of the vector).


If the user has a positive feedback on the item, we minimize the MSE between the event embedding $\textbf{e}_u^v$ and the $\mtrue$ vector, otherwise, if the user has a negative feedback on the item, we minimize the MSE between $\textbf{e}_u^v$ and the $\mfalse$ vector. The event embedding loss function is:
\begin{equation}
    \mathcal{L}_{event} = \sum_u\sum_{v\in\mathcal{V}_u^+}\text{MSE}(\textbf{e}_u^v, \textbf{G})
\end{equation}
where $\mathcal{V}_u^+$ is the set of user $u$'s interacted items, and $\textbf{G}$ represents the ground-truth vector, which is $\mtrue$ or $\mfalse$, depending on the user likes or dislikes the item. Then, we add this loss to the current loss function in Eq.\eqref{eq:model_los} to achieve the following new loss function:
\begin{equation}
    \mathcal{L}_{Boolean} = \mathcal{L}_{ncr} + \lambda_r\mathcal{L}_{logicReg} + \lambda_e\mathcal{L}_{event}
\end{equation}
where $\lambda_e$ is the coefficient of the event embedding loss. By adding this loss, the model tries to polarize the event embeddings to either $\mtrue$ or $\mfalse$. The model would then conduct reasoning in an approximate binary space when the event embedding loss $\mathcal{L}_{event}$ is minimized to a very small number (0.0001 in our experiment). We present the results of NDCG@10 on ML100k with $\lambda_{e}\in[0.001,0.01,0.1,1]$ in Figure~\ref{fig:logic_regularizer_weight}(d). Results on other datasets are similar.

From the results, we can see that the ranking performance heavily drops with the increase of the event embedding loss coefficient. A large $\lambda_e$ would limit the expressiveness power of the latent embeddings, which further limits the ability of logic neural networks to properly model the recommendation task. This experiment shows that it is important to blend the power of embedding learning into logical reasoning for accurate decision making in a logical space.


\section{Conclusions and Future Work}\label{sec:conclusion}
In this paper, we proposed a Neural Collaborative Reasoning (NCR) framework, which models recommendation as a reasoning task by integrating logical structures and neural networks for personalized recommendation. Experiments show that our method provides significant improvement on the ranking performance. We conducted further experiments to explore the behavior of our model under different settings, so as to understand why the model achieves good performance. Results show that appropriate logical regularization is helpful to the recommendation performance. 

Our work provides a very fundamental framework to integrate learning and reasoning, which conducts neural logic reasoning on top of learned vector representations. This inspires a wide scope of possibilities for future work. In this work, we only used the user interaction information for collaborative reasoning, while in the future, it is interesting to consider contextual and multimodal information for reasoning. Besides, the modularized design of our framework makes it promising to integrate with Neural Architecture Search (NAS) for Automatic Machine Learning (AutoML) \cite{elsken2019neural}, Program Synthesis \cite{gulwani2017program} and Explainable AI \cite{zhang2020explainable}. Finally, except for the recommendation task, the framework can also be extended to other tasks such as predicate logic reasoning, vision and language reasoning, knowledge graph reasoning, graph neural networks, social networks, and domain-specific applications such as medical and legal research where cognitive reasoning is significantly required, which are very promising directions to explore in the future.




\section*{Acknowledgement}
We thank the reviewers for the reviews and suggestions. This work was supported in part by NSF IIS-1910154 and IIS-2007907. Any opinions, findings, conclusions or recommendations expressed in this material are those of the authors and do not necessarily reflect those of the sponsors.

\appendix
\section*{APPENDIX}
\setcounter{section}{1}
\subsection{Data Preprocessing}\label{ap:datasets}
We transform the rating information, which comes with 1 to 5 ratings, to 0 and 1, which represents the negative or positive explicit feedback. Ratings equal to or higher than 4 are mapped to 1 (positive), while those equal to or lower than 3 are transformed to 0 (negative). Then we sort the dataset by timestamp. For each user and item pair in the dataset, we select its corresponding most recent $n$ history interactions to build the logical expression. Here we set the length of history to 5, which means that each user-item pair comes with 5 history interactions. For those items from the earliest 5 interactions of the corresponding user, we put them into the training dataset. Users with less than 5 interactions are put into the training dataset. We conduct leave-one-out operation to create the validation set and test set, which means that the last two interactions of each user are assigned to the validation set and the test set, respectively. Test sets are preferred if there remains only one expression for the user. For the models with implicit feedback as input, we simply ignore the rating information in the experiments.

\subsection{Pseudo-Code to Calculate the NCR Loss}\label{ap:algorithm}
The pseudo-code for calculating the NCR loss in Eq.\eqref{eq:l_ncr} is shown in Algorithm \ref{alg:la}.

\begin{algorithm}[t]
  \caption{Neural Collaborative Reasoning Loss (Eq.\eqref{eq:l_ncr})}\label{alg:la}
  \begin{flushleft}
 \textbf{Input:} Training user set $U$, item set $V$, model parameters $\Theta$, $\ell_2$-regularization coefficient $\lambda_\Theta$\\
 \textbf{Output:} Model loss 
 \end{flushleft}
  \begin{algorithmic}[1]
    \Procedure{CalcNCRLoss}{}
      \State Randomly initialize $\Theta$
      \State $Loss \gets 0$
      \For{$u\in U$}
        \State $V_{hist}\gets$ drawHistory$(u)$\Comment{obtain history of user $u$}
        \State $v_i\gets$drawTarget$(u)$\Comment{obtain the target item of user $u$}
        \State $V^- \gets$ Sample$(u, v_i, V_{hist}, V, n)$\Comment{get $n$ negative samples}
        \State $E\gets \text{ENCODE}(V_{hist})$\Comment{get history event embeddings}
        \State $\textbf{e}_i\gets \text{ENCODE}(v_i)$\Comment{get target event embedding}
        \State $s_{ui}^+\gets$ Sim$(LNN(E, \textbf{e}_i), \textbf{T})$\Comment{get target event score}
        \For{$v_j\in V^-$}
        \State $\textbf{e}_j\gets \text{ENCODE}(v_j)$\Comment{get negative event embedding}
        \State $s_{uj}^-\gets$ Sim$(LNN(E, \textbf{e}_j), \textbf{T})$\Comment{negative event score}
        \State $s_{uij}\gets \alpha\cdot(s_{ui}^+ - s_{uj}^-)$\Comment{get the final pair-wise score}
        \State $Loss \gets Loss + \ln{\sigma(s_{uij})}$\Comment{update loss}
        \EndFor
      \EndFor
      \State $Loss \gets -Loss + \lambda_\Theta\|\Theta\|_2^2$\Comment{update loss}
      \State \textbf{return} $Loss$\Comment{return loss}
    \EndProcedure
  \end{algorithmic}
\end{algorithm}

\subsection{Additional Experimental Settings}\label{ap:baseline}
We carefully tune the parameters for all baseline models to reach their best performance. For DMF, we implemented the model with two-layer neural networks to model users and items, respectively. Since the authors claimed that the increment of vector size would help to improve the performance, we tune the hidden vector size and set it to 128 to reach the best performance; For NeuMF, we use a three-layer multi-layer perceptron network with layer sizes 32, 16, 8 as mentioned by the author. The final output layer has only one layer with dimension 64; For both the GRU4Rec and STAMP models, we use 5 as the history length, which is the same as our model in the experiments. The size of hidden state vectors of both models are 64. For the CMPModel and EqModel, we apply the same parameters as our NCR model to guarantee that the differences in the reported results only result from the variation of the neural network structure.

For training process, early-stopping is conducted according to the performance on the validation set. Training examples that share the same neural network structure are put into the same mini-batch for better efficiency. We run the experiments with five different random seeds and report the best results of each model. The $p$-value of paired $t$-test are calculated based on these 5 round random experiments. We use both the $\ell_2$-regularization and dropout to prevent overfitting. The weight of $\ell_2$-regularization $\lambda_\Theta$ is set between $1\times10^{-6}$ to $1\times10^{-4}$ and dropout ratio is set to 0.2. Vector sizes of the variables and the user/item vectors are 64. The maximum training epoch is set to 100. 

All of the neural network parameters for DMF, NeuMF, GRU4Rec, STAMP, NLR, NCR, CMPModel and EqModel are initialized from a normal distribution with 0 mean and 0.01 standard deviation. Our framework is implemented with PyTorch~\cite{paszke2017automatic} v1.4 on an NVIDIA Geforce 2080Ti GPU. The operating system is Ubuntu 16.04 LTS.

%
%
\bibliographystyle{ACM-Reference-Format}
\balance
\bibliography{paper}


\begin{thebibliography}{58}


\ifx \showCODEN    \undefined \def \showCODEN     #1{\unskip}     \fi
\ifx \showDOI      \undefined \def \showDOI       #1{#1}\fi
\ifx \showISBNx    \undefined \def \showISBNx     #1{\unskip}     \fi
\ifx \showISBNxiii \undefined \def \showISBNxiii  #1{\unskip}     \fi
\ifx \showISSN     \undefined \def \showISSN      #1{\unskip}     \fi
\ifx \showLCCN     \undefined \def \showLCCN      #1{\unskip}     \fi
\ifx \shownote     \undefined \def \shownote      #1{#1}          \fi
\ifx \showarticletitle \undefined \def \showarticletitle #1{#1}   \fi
\ifx \showURL      \undefined \def \showURL       {\relax}        \fi
\providecommand\bibfield[2]{#2}
\providecommand\bibinfo[2]{#2}
\providecommand\natexlab[1]{#1}
\providecommand\showeprint[2][]{arXiv:#2}

\bibitem[\protect\citeauthoryear{Adomavicius, Mobasher, Ricci, and
  Tuzhilin}{Adomavicius et~al\mbox{.}}{2011}]%
        {cars}
\bibfield{author}{\bibinfo{person}{Gediminas Adomavicius},
  \bibinfo{person}{Bamshad Mobasher}, \bibinfo{person}{Francesco Ricci}, {and}
  \bibinfo{person}{Alex Tuzhilin}.} \bibinfo{year}{2011}\natexlab{}.
\newblock \showarticletitle{{Context-Aware Recommender Systems}}.
\newblock \bibinfo{journal}{\emph{Recommender systems handbook}}
  (\bibinfo{year}{2011}), \bibinfo{pages}{217--253}.
\newblock


\bibitem[\protect\citeauthoryear{Ai, Azizi, Chen, and Zhang}{Ai
  et~al\mbox{.}}{2018}]%
        {ai2018learning}
\bibfield{author}{\bibinfo{person}{Qingyao Ai}, \bibinfo{person}{Vahid Azizi},
  \bibinfo{person}{Xu Chen}, {and} \bibinfo{person}{Yongfeng Zhang}.}
  \bibinfo{year}{2018}\natexlab{}.
\newblock \showarticletitle{Learning heterogeneous knowledge base embeddings
  for explainable recommendation}.
\newblock \bibinfo{journal}{\emph{Algorithms}} \bibinfo{volume}{11},
  \bibinfo{number}{9} (\bibinfo{year}{2018}), \bibinfo{pages}{137}.
\newblock


\bibitem[\protect\citeauthoryear{Bellogin, Castells, and Cantador}{Bellogin
  et~al\mbox{.}}{2011}]%
        {bellogin2011precision}
\bibfield{author}{\bibinfo{person}{Alejandro Bellogin}, \bibinfo{person}{Pablo
  Castells}, {and} \bibinfo{person}{Ivan Cantador}.}
  \bibinfo{year}{2011}\natexlab{}.
\newblock \showarticletitle{Precision-oriented evaluation of recommender
  systems: an algorithmic comparison}. In \bibinfo{booktitle}{\emph{Proceedings
  of the fifth ACM conference on Recommender systems}}.
  \bibinfo{pages}{333--336}.
\newblock


\bibitem[\protect\citeauthoryear{Bengio}{Bengio}{2019}]%
        {bengio2019from}
\bibfield{author}{\bibinfo{person}{Yoshua Bengio}.}
  \bibinfo{year}{2019}\natexlab{}.
\newblock \showarticletitle{From System 1 Deep Learning to System 2 Deep
  Learning}. In \bibinfo{booktitle}{\emph{NeurIPS'2019}}.
\newblock


\bibitem[\protect\citeauthoryear{Besold, Garcez, Bader, Bowman, Domingos,
  Hitzler, K{\"u}hnberger, Lamb, Lowd, Lima, et~al\mbox{.}}{Besold
  et~al\mbox{.}}{2017}]%
        {besold2017neural}
\bibfield{author}{\bibinfo{person}{Tarek~R Besold},
  \bibinfo{person}{Artur~d'Avila Garcez}, \bibinfo{person}{Sebastian Bader},
  \bibinfo{person}{Howard Bowman}, \bibinfo{person}{Pedro Domingos},
  \bibinfo{person}{Pascal Hitzler}, \bibinfo{person}{Kai-Uwe K{\"u}hnberger},
  \bibinfo{person}{Luis~C Lamb}, \bibinfo{person}{Daniel Lowd},
  \bibinfo{person}{Priscila Machado~Vieira Lima}, {et~al\mbox{.}}}
  \bibinfo{year}{2017}\natexlab{}.
\newblock \showarticletitle{Neural-symbolic learning and reasoning: A survey
  and interpretation}.
\newblock \bibinfo{journal}{\emph{arXiv preprint arXiv:1711.03902}}
  (\bibinfo{year}{2017}).
\newblock


\bibitem[\protect\citeauthoryear{Chen, Xu, Zhang, Tang, Cao, Qin, and Zha}{Chen
  et~al\mbox{.}}{2018}]%
        {chen2018sequential}
\bibfield{author}{\bibinfo{person}{Xu Chen}, \bibinfo{person}{Hongteng Xu},
  \bibinfo{person}{Yongfeng Zhang}, \bibinfo{person}{Jiaxi Tang},
  \bibinfo{person}{Yixin Cao}, \bibinfo{person}{Zheng Qin}, {and}
  \bibinfo{person}{Hongyuan Zha}.} \bibinfo{year}{2018}\natexlab{}.
\newblock \showarticletitle{Sequential recommendation with user memory
  networks}. In \bibinfo{booktitle}{\emph{WSDM}}. \bibinfo{pages}{108--116}.
\newblock


\bibitem[\protect\citeauthoryear{Cheng, Koc, Harmsen, Shaked, Chandra, Aradhye,
  Anderson, Corrado, Chai, Ispir, et~al\mbox{.}}{Cheng et~al\mbox{.}}{2016}]%
        {cheng2016wide}
\bibfield{author}{\bibinfo{person}{Heng-Tze Cheng}, \bibinfo{person}{Levent
  Koc}, \bibinfo{person}{Jeremiah Harmsen}, \bibinfo{person}{Tal Shaked},
  \bibinfo{person}{Tushar Chandra}, \bibinfo{person}{Hrishi Aradhye},
  \bibinfo{person}{Glen Anderson}, \bibinfo{person}{Greg Corrado},
  \bibinfo{person}{Wei Chai}, \bibinfo{person}{Mustafa Ispir}, {et~al\mbox{.}}}
  \bibinfo{year}{2016}\natexlab{}.
\newblock \showarticletitle{Wide \& deep learning for recommender systems}. In
  \bibinfo{booktitle}{\emph{DLRS: Proceedings of the 1st RecSys workshop on
  deep learning for recommender systems}}. \bibinfo{pages}{7--10}.
\newblock


\bibitem[\protect\citeauthoryear{Dacrema, Boglio, Cremonesi, and
  Jannach}{Dacrema et~al\mbox{.}}{2021}]%
        {dacrema2021troubling}
\bibfield{author}{\bibinfo{person}{Maurizio~Ferrari Dacrema},
  \bibinfo{person}{Simone Boglio}, \bibinfo{person}{Paolo Cremonesi}, {and}
  \bibinfo{person}{Dietmar Jannach}.} \bibinfo{year}{2021}\natexlab{}.
\newblock \showarticletitle{A troubling analysis of reproducibility and
  progress in recommender systems research}.
\newblock \bibinfo{journal}{\emph{ACM Transactions on Information Systems
  (TOIS)}} (\bibinfo{year}{2021}).
\newblock


\bibitem[\protect\citeauthoryear{Dacrema, Cremonesi, and Jannach}{Dacrema
  et~al\mbox{.}}{2019}]%
        {dacrema2019we}
\bibfield{author}{\bibinfo{person}{Maurizio~Ferrari Dacrema},
  \bibinfo{person}{Paolo Cremonesi}, {and} \bibinfo{person}{Dietmar Jannach}.}
  \bibinfo{year}{2019}\natexlab{}.
\newblock \showarticletitle{Are we really making much progress? A worrying
  analysis of recent neural recommendation approaches}. In
  \bibinfo{booktitle}{\emph{RecSys}}. \bibinfo{pages}{101--109}.
\newblock


\bibitem[\protect\citeauthoryear{Dai, Xu, Yu, and Zhou}{Dai
  et~al\mbox{.}}{2019}]%
        {dai2019bridging}
\bibfield{author}{\bibinfo{person}{Wang-Zhou Dai}, \bibinfo{person}{Qiuling
  Xu}, \bibinfo{person}{Yang Yu}, {and} \bibinfo{person}{Zhi-Hua Zhou}.}
  \bibinfo{year}{2019}\natexlab{}.
\newblock \showarticletitle{Bridging machine learning and logical reasoning by
  abductive learning}. In \bibinfo{booktitle}{\emph{NeurIPS}}.
  \bibinfo{pages}{2815--2826}.
\newblock


\bibitem[\protect\citeauthoryear{Dong, Mao, Lin, Wang, Li, and Zhou}{Dong
  et~al\mbox{.}}{2019}]%
        {dong2019neural}
\bibfield{author}{\bibinfo{person}{Honghua Dong}, \bibinfo{person}{Jiayuan
  Mao}, \bibinfo{person}{Tian Lin}, \bibinfo{person}{Chong Wang},
  \bibinfo{person}{Lihong Li}, {and} \bibinfo{person}{Denny Zhou}.}
  \bibinfo{year}{2019}\natexlab{}.
\newblock \showarticletitle{Neural logic machines}.
\newblock \bibinfo{journal}{\emph{ICLR}}.
\newblock


\bibitem[\protect\citeauthoryear{Ekstrand, Riedl, Konstan,
  et~al\mbox{.}}{Ekstrand et~al\mbox{.}}{2011}]%
        {ekstrand2011collaborative}
\bibfield{author}{\bibinfo{person}{Michael~D Ekstrand}, \bibinfo{person}{John~T
  Riedl}, \bibinfo{person}{Joseph~A Konstan}, {et~al\mbox{.}}}
  \bibinfo{year}{2011}\natexlab{}.
\newblock \showarticletitle{Collaborative filtering recommender systems}.
\newblock \bibinfo{journal}{\emph{Foundations and Trends{\textregistered} in
  Human--Computer Interaction}} \bibinfo{volume}{4}, \bibinfo{number}{2}
  (\bibinfo{year}{2011}), \bibinfo{pages}{81--173}.
\newblock


\bibitem[\protect\citeauthoryear{Elsken, Metzen, Hutter, et~al\mbox{.}}{Elsken
  et~al\mbox{.}}{2019}]%
        {elsken2019neural}
\bibfield{author}{\bibinfo{person}{Thomas Elsken}, \bibinfo{person}{Jan~Hendrik
  Metzen}, \bibinfo{person}{Frank Hutter}, {et~al\mbox{.}}}
  \bibinfo{year}{2019}\natexlab{}.
\newblock \showarticletitle{Neural architecture search: A survey.}
\newblock \bibinfo{journal}{\emph{Journal of Machine Learning Research}}
  \bibinfo{volume}{20}, \bibinfo{number}{55} (\bibinfo{year}{2019}),
  \bibinfo{pages}{1--21}.
\newblock


\bibitem[\protect\citeauthoryear{Ferrari~Dacrema, Parroni, Cremonesi, and
  Jannach}{Ferrari~Dacrema et~al\mbox{.}}{2020}]%
        {ferrari2020critically}
\bibfield{author}{\bibinfo{person}{Maurizio Ferrari~Dacrema},
  \bibinfo{person}{Federico Parroni}, \bibinfo{person}{Paolo Cremonesi}, {and}
  \bibinfo{person}{Dietmar Jannach}.} \bibinfo{year}{2020}\natexlab{}.
\newblock \showarticletitle{Critically Examining the Claimed Value of
  Convolutions over User-Item Embedding Maps for Recommender Systems}. In
  \bibinfo{booktitle}{\emph{CIKM}}. \bibinfo{pages}{355--363}.
\newblock


\bibitem[\protect\citeauthoryear{Garcez, Broda, and Gabbay}{Garcez
  et~al\mbox{.}}{2012}]%
        {garcez2012neural}
\bibfield{author}{\bibinfo{person}{Artur S~d'Avila Garcez},
  \bibinfo{person}{Krysia~B Broda}, {and} \bibinfo{person}{Dov~M Gabbay}.}
  \bibinfo{year}{2012}\natexlab{}.
\newblock \bibinfo{booktitle}{\emph{Neural-symbolic learning systems:
  foundations and applications}}.
\newblock \bibinfo{publisher}{Springer Sci. \& Bus. Media}.
\newblock


\bibitem[\protect\citeauthoryear{Gulwani, Polozov, Singh,
  et~al\mbox{.}}{Gulwani et~al\mbox{.}}{2017}]%
        {gulwani2017program}
\bibfield{author}{\bibinfo{person}{Sumit Gulwani}, \bibinfo{person}{Oleksandr
  Polozov}, \bibinfo{person}{Rishabh Singh}, {et~al\mbox{.}}}
  \bibinfo{year}{2017}\natexlab{}.
\newblock \showarticletitle{Program synthesis}.
\newblock \bibinfo{journal}{\emph{Foundations and Trends{\textregistered} in
  Programming Languages}} \bibinfo{volume}{4}, \bibinfo{number}{1-2}
  (\bibinfo{year}{2017}), \bibinfo{pages}{1--119}.
\newblock


\bibitem[\protect\citeauthoryear{Harper and Konstan}{Harper and
  Konstan}{2016}]%
        {harper2016movielens}
\bibfield{author}{\bibinfo{person}{F~Maxwell Harper} {and}
  \bibinfo{person}{Joseph~A Konstan}.} \bibinfo{year}{2016}\natexlab{}.
\newblock \showarticletitle{The movielens datasets: History and context}.
\newblock \bibinfo{journal}{\emph{Acm TIST}} \bibinfo{volume}{5},
  \bibinfo{number}{4} (\bibinfo{year}{2016}), \bibinfo{pages}{19}.
\newblock


\bibitem[\protect\citeauthoryear{He, Kang, and McAuley}{He
  et~al\mbox{.}}{2017a}]%
        {he2017translation}
\bibfield{author}{\bibinfo{person}{Ruining He}, \bibinfo{person}{Wang-Cheng
  Kang}, {and} \bibinfo{person}{Julian McAuley}.}
  \bibinfo{year}{2017}\natexlab{a}.
\newblock \showarticletitle{Translation-based recommendation}. In
  \bibinfo{booktitle}{\emph{RecSys}}. \bibinfo{pages}{161--169}.
\newblock


\bibitem[\protect\citeauthoryear{He and McAuley}{He and McAuley}{2016}]%
        {he2016vbpr}
\bibfield{author}{\bibinfo{person}{Ruining He} {and} \bibinfo{person}{Julian
  McAuley}.} \bibinfo{year}{2016}\natexlab{}.
\newblock \showarticletitle{VBPR: Visual Bayesian Personalized Ranking from
  Implicit Feedback}. In \bibinfo{booktitle}{\emph{AAAI}}.
\newblock


\bibitem[\protect\citeauthoryear{He, Liao, Zhang, Nie, Hu, and Chua}{He
  et~al\mbox{.}}{2017b}]%
        {he2017neural}
\bibfield{author}{\bibinfo{person}{Xiangnan He}, \bibinfo{person}{Lizi Liao},
  \bibinfo{person}{Hanwang Zhang}, \bibinfo{person}{Liqiang Nie},
  \bibinfo{person}{Xia Hu}, {and} \bibinfo{person}{Tat-Seng Chua}.}
  \bibinfo{year}{2017}\natexlab{b}.
\newblock \showarticletitle{Neural Collaborative Filtering}. In
  \bibinfo{booktitle}{\emph{WWW}}. \bibinfo{pages}{173--182}.
\newblock


\bibitem[\protect\citeauthoryear{Hidasi, Karatzoglou, Baltrunas, and
  Tikk}{Hidasi et~al\mbox{.}}{2016}]%
        {hidasi2016session}
\bibfield{author}{\bibinfo{person}{Bal{\'a}zs Hidasi},
  \bibinfo{person}{Alexandros Karatzoglou}, \bibinfo{person}{Linas Baltrunas},
  {and} \bibinfo{person}{D Tikk}.} \bibinfo{year}{2016}\natexlab{}.
\newblock \showarticletitle{Session-based recommendations with recurrent neural
  networks}. In \bibinfo{booktitle}{\emph{ICLR}}.
\newblock


\bibitem[\protect\citeauthoryear{Hsieh, Yang, Cui, Lin, Belongie, and
  Estrin}{Hsieh et~al\mbox{.}}{2017}]%
        {hsieh2017collaborative}
\bibfield{author}{\bibinfo{person}{Cheng-Kang Hsieh}, \bibinfo{person}{Longqi
  Yang}, \bibinfo{person}{Yin Cui}, \bibinfo{person}{Tsung-Yi Lin},
  \bibinfo{person}{Serge Belongie}, {and} \bibinfo{person}{Deborah Estrin}.}
  \bibinfo{year}{2017}\natexlab{}.
\newblock \showarticletitle{Collaborative metric learning}. In
  \bibinfo{booktitle}{\emph{In WWW}}. \bibinfo{pages}{193--201}.
\newblock


\bibitem[\protect\citeauthoryear{Jiang and Luo}{Jiang and Luo}{2019}]%
        {jiang2019neural}
\bibfield{author}{\bibinfo{person}{Zhengyao Jiang} {and} \bibinfo{person}{Shan
  Luo}.} \bibinfo{year}{2019}\natexlab{}.
\newblock \showarticletitle{Neural Logic Reinforcement Learning}.
\newblock \bibinfo{journal}{\emph{Proceedings of the 36th International
  Conference on Machine Learning}} (\bibinfo{year}{2019}).
\newblock


\bibitem[\protect\citeauthoryear{Kang and McAuley}{Kang and McAuley}{2018}]%
        {kang2018self}
\bibfield{author}{\bibinfo{person}{Wang-Cheng Kang} {and}
  \bibinfo{person}{Julian McAuley}.} \bibinfo{year}{2018}\natexlab{}.
\newblock \showarticletitle{Self-attentive sequential recommendation}. In
  \bibinfo{booktitle}{\emph{2018 IEEE International Conference on Data Mining
  (ICDM)}}. IEEE.
\newblock


\bibitem[\protect\citeauthoryear{Karatzoglou, Amatriain, Baltrunas, and
  Oliver}{Karatzoglou et~al\mbox{.}}{2010}]%
        {multiverse}
\bibfield{author}{\bibinfo{person}{Alexandros Karatzoglou},
  \bibinfo{person}{Xavier Amatriain}, \bibinfo{person}{Linas Baltrunas}, {and}
  \bibinfo{person}{Nuria Oliver}.} \bibinfo{year}{2010}\natexlab{}.
\newblock \showarticletitle{{Multiverse Recommendation: N-dimensional Tensor
  Factorization for Context-aware Collaborative Filtering}}.
\newblock \bibinfo{journal}{\emph{RecSys}} (\bibinfo{year}{2010}),
  \bibinfo{pages}{79--86}.
\newblock


\bibitem[\protect\citeauthoryear{Kingma and Ba}{Kingma and Ba}{2014}]%
        {kingma2014adam}
\bibfield{author}{\bibinfo{person}{Diederik~P Kingma} {and}
  \bibinfo{person}{Jimmy Ba}.} \bibinfo{year}{2014}\natexlab{}.
\newblock \showarticletitle{Adam: A method for stochastic optimization}.
\newblock \bibinfo{journal}{\emph{arXiv preprint arXiv:1412.6980}}
  (\bibinfo{year}{2014}).
\newblock


\bibitem[\protect\citeauthoryear{Konstan, Miller, Maltz, Herlocker, Gordon, and
  Riedl}{Konstan et~al\mbox{.}}{1997}]%
        {konstan1997grouplens}
\bibfield{author}{\bibinfo{person}{Joseph~A Konstan},
  \bibinfo{person}{Bradley~N Miller}, \bibinfo{person}{David Maltz},
  \bibinfo{person}{Jonathan~L Herlocker}, \bibinfo{person}{Lee~R Gordon}, {and}
  \bibinfo{person}{John Riedl}.} \bibinfo{year}{1997}\natexlab{}.
\newblock \showarticletitle{GroupLens: applying collaborative filtering to
  Usenet news}.
\newblock \bibinfo{journal}{\emph{Commun. ACM}} \bibinfo{volume}{40},
  \bibinfo{number}{3} (\bibinfo{year}{1997}), \bibinfo{pages}{77--87}.
\newblock


\bibitem[\protect\citeauthoryear{Koren}{Koren}{2008}]%
        {koren2008factorization}
\bibfield{author}{\bibinfo{person}{Yehuda Koren}.}
  \bibinfo{year}{2008}\natexlab{}.
\newblock \showarticletitle{Factorization meets the neighborhood: a
  multifaceted collaborative filtering model}. In
  \bibinfo{booktitle}{\emph{SIGKDD}}. ACM, \bibinfo{pages}{426--434}.
\newblock


\bibitem[\protect\citeauthoryear{Koren}{Koren}{2009}]%
        {cf-temporal}
\bibfield{author}{\bibinfo{person}{Yehuda Koren}.}
  \bibinfo{year}{2009}\natexlab{}.
\newblock \showarticletitle{{Collaborative filtering with temporal dynamics}}.
\newblock \bibinfo{journal}{\emph{KDD}} (\bibinfo{year}{2009}).
\newblock


\bibitem[\protect\citeauthoryear{Koren, Bell, and Volinsky}{Koren
  et~al\mbox{.}}{2009}]%
        {koren2009matrix}
\bibfield{author}{\bibinfo{person}{Yehuda Koren}, \bibinfo{person}{Robert
  Bell}, {and} \bibinfo{person}{Chris Volinsky}.}
  \bibinfo{year}{2009}\natexlab{}.
\newblock \showarticletitle{Matrix factorization techniques for recommender
  systems}.
\newblock \bibinfo{journal}{\emph{Computer}} \bibinfo{number}{8}
  (\bibinfo{year}{2009}), \bibinfo{pages}{30--37}.
\newblock


\bibitem[\protect\citeauthoryear{Lee and Seung}{Lee and Seung}{2001}]%
        {lee2001algorithms}
\bibfield{author}{\bibinfo{person}{Daniel~D Lee} {and}
  \bibinfo{person}{H~Sebastian Seung}.} \bibinfo{year}{2001}\natexlab{}.
\newblock \showarticletitle{Algorithms for non-negative matrix factorization}.
  In \bibinfo{booktitle}{\emph{Advances in neural information processing
  systems}}. \bibinfo{pages}{556--562}.
\newblock


\bibitem[\protect\citeauthoryear{Li, Ren, Chen, Ren, Lian, and Ma}{Li
  et~al\mbox{.}}{2017}]%
        {li2017neural}
\bibfield{author}{\bibinfo{person}{Jing Li}, \bibinfo{person}{Pengjie Ren},
  \bibinfo{person}{Zhumin Chen}, \bibinfo{person}{Zhaochun Ren},
  \bibinfo{person}{Tao Lian}, {and} \bibinfo{person}{Jun Ma}.}
  \bibinfo{year}{2017}\natexlab{}.
\newblock \showarticletitle{Neural attentive session-based recommendation}. In
  \bibinfo{booktitle}{\emph{CIKM}}. \bibinfo{pages}{1419--1428}.
\newblock


\bibitem[\protect\citeauthoryear{Linden, Smith, and York}{Linden
  et~al\mbox{.}}{2003}]%
        {Linden2003}
\bibfield{author}{\bibinfo{person}{Greg Linden}, \bibinfo{person}{Brent Smith},
  {and} \bibinfo{person}{Jeremy York}.} \bibinfo{year}{2003}\natexlab{}.
\newblock \showarticletitle{Amazon.Com Recommendations: Item-to-Item
  Collaborative Filtering}.
\newblock \bibinfo{journal}{\emph{IEEE Internet Computing}}
  (\bibinfo{year}{2003}).
\newblock


\bibitem[\protect\citeauthoryear{Marcus}{Marcus}{2020}]%
        {marcus2020next}
\bibfield{author}{\bibinfo{person}{Gary Marcus}.}
  \bibinfo{year}{2020}\natexlab{}.
\newblock \showarticletitle{The next decade in ai: four steps towards robust
  artificial intelligence}.
\newblock \bibinfo{journal}{\emph{arXiv preprint arXiv:2002.06177}}
  (\bibinfo{year}{2020}).
\newblock


\bibitem[\protect\citeauthoryear{McAuley, Targett, Shi, and van~den
  Hengel}{McAuley et~al\mbox{.}}{2015}]%
        {mcauley2015image}
\bibfield{author}{\bibinfo{person}{Julian McAuley},
  \bibinfo{person}{Christopher Targett}, \bibinfo{person}{Qinfeng Shi}, {and}
  \bibinfo{person}{Anton van~den Hengel}.} \bibinfo{year}{2015}\natexlab{}.
\newblock \showarticletitle{Image-based recommendations on styles and
  substitutes}. In \bibinfo{booktitle}{\emph{SIGIR}}. ACM.
\newblock


\bibitem[\protect\citeauthoryear{McCelloch and Pitts}{McCelloch and
  Pitts}{1943}]%
        {mccelloch1943logical}
\bibfield{author}{\bibinfo{person}{WS McCelloch} {and} \bibinfo{person}{Walter
  Pitts}.} \bibinfo{year}{1943}\natexlab{}.
\newblock \showarticletitle{A Logical Calculus of the Idea Immanent in Neural
  Nets}.
\newblock \bibinfo{journal}{\emph{Bulletin ofMathematical Biophysics}}
  \bibinfo{volume}{5} (\bibinfo{year}{1943}), \bibinfo{pages}{115--133}.
\newblock


\bibitem[\protect\citeauthoryear{Mikolov, Sutskever, Chen, Corrado, and
  Dean}{Mikolov et~al\mbox{.}}{2013}]%
        {mikolov2013distributed}
\bibfield{author}{\bibinfo{person}{Tomas Mikolov}, \bibinfo{person}{Ilya
  Sutskever}, \bibinfo{person}{Kai Chen}, \bibinfo{person}{Greg~S Corrado},
  {and} \bibinfo{person}{Jeff Dean}.} \bibinfo{year}{2013}\natexlab{}.
\newblock \showarticletitle{Distributed representations of words and phrases
  and their compositionality}. In \bibinfo{booktitle}{\emph{NeurIPS}}.
  \bibinfo{pages}{3111--3119}.
\newblock


\bibitem[\protect\citeauthoryear{Mnih and Salakhutdinov}{Mnih and
  Salakhutdinov}{2008}]%
        {mnih2008probabilistic}
\bibfield{author}{\bibinfo{person}{Andriy Mnih} {and} \bibinfo{person}{Ruslan~R
  Salakhutdinov}.} \bibinfo{year}{2008}\natexlab{}.
\newblock \showarticletitle{Probabilistic matrix factorization}. In
  \bibinfo{booktitle}{\emph{Advances in neural information processing
  systems}}. \bibinfo{pages}{1257--1264}.
\newblock


\bibitem[\protect\citeauthoryear{Paszke, Gross, Chintala, Chanan, Yang, DeVito,
  Lin, Desmaison, Antiga, and Lerer}{Paszke et~al\mbox{.}}{2017}]%
        {paszke2017automatic}
\bibfield{author}{\bibinfo{person}{Adam Paszke}, \bibinfo{person}{Sam Gross},
  \bibinfo{person}{Soumith Chintala}, \bibinfo{person}{Gregory Chanan},
  \bibinfo{person}{Edward Yang}, \bibinfo{person}{Zachary DeVito},
  \bibinfo{person}{Zeming Lin}, \bibinfo{person}{Alban Desmaison},
  \bibinfo{person}{Luca Antiga}, {and} \bibinfo{person}{Adam Lerer}.}
  \bibinfo{year}{2017}\natexlab{}.
\newblock \showarticletitle{Automatic differentiation in PyTorch}.
\newblock  (\bibinfo{year}{2017}).
\newblock


\bibitem[\protect\citeauthoryear{{Qiao Liu, Yifu Zeng, Refuoe Mokhosi, Haibin
  Zhang}}{{Qiao Liu, Yifu Zeng, Refuoe Mokhosi, Haibin Zhang}}{2018}]%
        {liu2018stamp}
\bibfield{author}{\bibinfo{person}{{Qiao Liu, Yifu Zeng, Refuoe Mokhosi, Haibin
  Zhang}}.} \bibinfo{year}{2018}\natexlab{}.
\newblock \showarticletitle{STAMP: short-term attention/memory priority model
  for session-based recommendation}. In \bibinfo{booktitle}{\emph{KDD}}.
\newblock


\bibitem[\protect\citeauthoryear{Qu and Tang}{Qu and Tang}{2019}]%
        {qu2019probabilistic}
\bibfield{author}{\bibinfo{person}{Meng Qu} {and} \bibinfo{person}{Jian Tang}.}
  \bibinfo{year}{2019}\natexlab{}.
\newblock \showarticletitle{Probabilistic logic neural networks for reasoning}.
  In \bibinfo{booktitle}{\emph{Advances in Neural Information Processing
  Systems}}. \bibinfo{pages}{7710--7720}.
\newblock


\bibitem[\protect\citeauthoryear{Rendle, Freudenthaler, Gantner, and
  Schmidt-Thieme}{Rendle et~al\mbox{.}}{2009}]%
        {bpr}
\bibfield{author}{\bibinfo{person}{Steffen Rendle}, \bibinfo{person}{Christoph
  Freudenthaler}, \bibinfo{person}{Zeno Gantner}, {and} \bibinfo{person}{Lars
  Schmidt-Thieme}.} \bibinfo{year}{2009}\natexlab{}.
\newblock \showarticletitle{BPR: Bayesian personalized ranking from implicit
  feedback}. In \bibinfo{booktitle}{\emph{UAI}}.
\newblock


\bibitem[\protect\citeauthoryear{Rendle, Krichene, Zhang, and Anderson}{Rendle
  et~al\mbox{.}}{2020}]%
        {rendle2020neural}
\bibfield{author}{\bibinfo{person}{Steffen Rendle}, \bibinfo{person}{Walid
  Krichene}, \bibinfo{person}{Li Zhang}, {and} \bibinfo{person}{John
  Anderson}.} \bibinfo{year}{2020}\natexlab{}.
\newblock \showarticletitle{Neural Collaborative Filtering vs. Matrix
  Factorization Revisited}.
\newblock \bibinfo{journal}{\emph{RecSys}} (\bibinfo{year}{2020}).
\newblock


\bibitem[\protect\citeauthoryear{Resnick, Iacovou, Suchak, Bergstrom, and
  Riedl}{Resnick et~al\mbox{.}}{1994}]%
        {resnick1994grouplens}
\bibfield{author}{\bibinfo{person}{Paul Resnick}, \bibinfo{person}{Neophytos
  Iacovou}, \bibinfo{person}{Mitesh Suchak}, \bibinfo{person}{Peter Bergstrom},
  {and} \bibinfo{person}{John Riedl}.} \bibinfo{year}{1994}\natexlab{}.
\newblock \showarticletitle{GroupLens: an open architecture for collaborative
  filtering of netnews}. In \bibinfo{booktitle}{\emph{CSCW}}. ACM,
  \bibinfo{pages}{175--186}.
\newblock


\bibitem[\protect\citeauthoryear{Ricci, Rokach, and Shapira}{Ricci
  et~al\mbox{.}}{2011}]%
        {rs}
\bibfield{author}{\bibinfo{person}{Francesco Ricci}, \bibinfo{person}{Lior
  Rokach}, {and} \bibinfo{person}{Bracha Shapira}.}
  \bibinfo{year}{2011}\natexlab{}.
\newblock \showarticletitle{{Introduction to Recommender Systems Handbook}}.
\newblock \bibinfo{journal}{\emph{Springer US}} (\bibinfo{year}{2011}).
\newblock


\bibitem[\protect\citeauthoryear{Richardson and Domingos}{Richardson and
  Domingos}{2006}]%
        {richardson2006markov}
\bibfield{author}{\bibinfo{person}{Matthew Richardson} {and}
  \bibinfo{person}{Pedro Domingos}.} \bibinfo{year}{2006}\natexlab{}.
\newblock \showarticletitle{Markov logic networks}.
\newblock \bibinfo{journal}{\emph{Machine learning}} \bibinfo{volume}{62},
  \bibinfo{number}{1-2} (\bibinfo{year}{2006}), \bibinfo{pages}{107--136}.
\newblock


\bibitem[\protect\citeauthoryear{Rumelhart, Hinton, and Williams}{Rumelhart
  et~al\mbox{.}}{1986}]%
        {rumelhart1986learning}
\bibfield{author}{\bibinfo{person}{David~E Rumelhart},
  \bibinfo{person}{Geoffrey~E Hinton}, {and} \bibinfo{person}{Ronald~J
  Williams}.} \bibinfo{year}{1986}\natexlab{}.
\newblock \showarticletitle{Learning representations by back-propagating
  errors}.
\newblock \bibinfo{journal}{\emph{nature}} \bibinfo{volume}{323},
  \bibinfo{number}{6088} (\bibinfo{year}{1986}), \bibinfo{pages}{533--536}.
\newblock


\bibitem[\protect\citeauthoryear{Said and Bellog{\'\i}n}{Said and
  Bellog{\'\i}n}{2014}]%
        {said2014comparative}
\bibfield{author}{\bibinfo{person}{Alan Said} {and} \bibinfo{person}{Alejandro
  Bellog{\'\i}n}.} \bibinfo{year}{2014}\natexlab{}.
\newblock \showarticletitle{Comparative recommender system evaluation:
  benchmarking recommendation frameworks}. In
  \bibinfo{booktitle}{\emph{RecSys}}.
\newblock


\bibitem[\protect\citeauthoryear{Sarwar, Karypis, Konstan, and Riedl}{Sarwar
  et~al\mbox{.}}{2001}]%
        {sarwar2001item}
\bibfield{author}{\bibinfo{person}{Badrul Sarwar}, \bibinfo{person}{George
  Karypis}, \bibinfo{person}{Joseph Konstan}, {and} \bibinfo{person}{John
  Riedl}.} \bibinfo{year}{2001}\natexlab{}.
\newblock \showarticletitle{Item-based collaborative filtering recommendation
  algorithms}. In \bibinfo{booktitle}{\emph{WWW}}. ACM,
  \bibinfo{pages}{285--295}.
\newblock


\bibitem[\protect\citeauthoryear{Shi, Chen, Ma, Mao, Zhang, and Zhang}{Shi
  et~al\mbox{.}}{2020}]%
        {shi2020neural}
\bibfield{author}{\bibinfo{person}{Shaoyun Shi}, \bibinfo{person}{Hanxiong
  Chen}, \bibinfo{person}{Weizhi Ma}, \bibinfo{person}{Jiaxin Mao},
  \bibinfo{person}{Min Zhang}, {and} \bibinfo{person}{Yongfeng Zhang}.}
  \bibinfo{year}{2020}\natexlab{}.
\newblock \showarticletitle{Neural Logic Reasoning}. In
  \bibinfo{booktitle}{\emph{CIKM}}. ACM.
\newblock


\bibitem[\protect\citeauthoryear{Xue, Dai, Zhang, Huang, and Chen}{Xue
  et~al\mbox{.}}{2017}]%
        {xue2017deep}
\bibfield{author}{\bibinfo{person}{Hong-Jian Xue}, \bibinfo{person}{Xinyu Dai},
  \bibinfo{person}{Jianbing Zhang}, \bibinfo{person}{Shujian Huang}, {and}
  \bibinfo{person}{Jiajun Chen}.} \bibinfo{year}{2017}\natexlab{}.
\newblock \showarticletitle{Deep Matrix Factorization Models for Recommender
  Systems.}. In \bibinfo{booktitle}{\emph{IJCAI}}.
\newblock


\bibitem[\protect\citeauthoryear{Zhang, Yuan, Lian, Xie, and Ma}{Zhang
  et~al\mbox{.}}{2016}]%
        {zhang2016collaborativekdd}
\bibfield{author}{\bibinfo{person}{Fuzheng Zhang},
  \bibinfo{person}{Nicholas~Jing Yuan}, \bibinfo{person}{Defu Lian},
  \bibinfo{person}{Xing Xie}, {and} \bibinfo{person}{Wei-Ying Ma}.}
  \bibinfo{year}{2016}\natexlab{}.
\newblock \showarticletitle{Collaborative Knowledge Base Embedding for
  Recommender Systems}. In \bibinfo{booktitle}{\emph{KDD}}.
\newblock


\bibitem[\protect\citeauthoryear{Zhang, Yao, and Sun}{Zhang
  et~al\mbox{.}}{2019}]%
        {zhang2019deep}
\bibfield{author}{\bibinfo{person}{Shuai Zhang}, \bibinfo{person}{Lina Yao},
  {and} \bibinfo{person}{Aixin Sun}.} \bibinfo{year}{2019}\natexlab{}.
\newblock \showarticletitle{Deep learning based recommender system: A survey
  and new perspectives}.
\newblock \bibinfo{journal}{\emph{Comput. Surveys}} \bibinfo{volume}{52},
  \bibinfo{number}{1} (\bibinfo{year}{2019}), \bibinfo{pages}{1--38}.
\newblock


\bibitem[\protect\citeauthoryear{Zhang, Ai, Chen, and Croft}{Zhang
  et~al\mbox{.}}{2017}]%
        {zhang2017joint}
\bibfield{author}{\bibinfo{person}{Yongfeng Zhang}, \bibinfo{person}{Qingyao
  Ai}, \bibinfo{person}{Xu Chen}, {and} \bibinfo{person}{W~Bruce Croft}.}
  \bibinfo{year}{2017}\natexlab{}.
\newblock \showarticletitle{Joint representation learning for top-n
  recommendation with heterogeneous information sources}. In
  \bibinfo{booktitle}{\emph{CIKM}}. ACM, \bibinfo{pages}{1449--1458}.
\newblock


\bibitem[\protect\citeauthoryear{Zhang and Chen}{Zhang and Chen}{2020}]%
        {zhang2020explainable}
\bibfield{author}{\bibinfo{person}{Yongfeng Zhang} {and} \bibinfo{person}{Xu
  Chen}.} \bibinfo{year}{2020}\natexlab{}.
\newblock \showarticletitle{Explainable recommendation: A survey and new
  perspectives}.
\newblock \bibinfo{journal}{\emph{Foundations and Trends in Information
  Retrieval}} \bibinfo{volume}{14}, \bibinfo{number}{1} (\bibinfo{year}{2020}).
\newblock


\bibitem[\protect\citeauthoryear{Zhang, Chen, Yang, Ramamurthy, Li, Qi, and
  Song}{Zhang et~al\mbox{.}}{2020}]%
        {zhang2020efficient}
\bibfield{author}{\bibinfo{person}{Yuyu Zhang}, \bibinfo{person}{Xinshi Chen},
  \bibinfo{person}{Yuan Yang}, \bibinfo{person}{Arun Ramamurthy},
  \bibinfo{person}{Bo Li}, \bibinfo{person}{Yuan Qi}, {and} \bibinfo{person}{Le
  Song}.} \bibinfo{year}{2020}\natexlab{}.
\newblock \showarticletitle{Efficient probabilistic logic reasoning with graph
  neural networks}.
\newblock \bibinfo{journal}{\emph{ICLR}} (\bibinfo{year}{2020}).
\newblock


\bibitem[\protect\citeauthoryear{Zhao, Chen, Wang, Gu, and Wen}{Zhao
  et~al\mbox{.}}{2020}]%
        {zhao2020revisiting}
\bibfield{author}{\bibinfo{person}{Wayne~Xin Zhao}, \bibinfo{person}{Junhua
  Chen}, \bibinfo{person}{Pengfei Wang}, \bibinfo{person}{Qi Gu}, {and}
  \bibinfo{person}{Ji-Rong Wen}.} \bibinfo{year}{2020}\natexlab{}.
\newblock \showarticletitle{Revisiting Alternative Experimental Settings for
  Evaluating Top-N Item Recommendation Algorithms}.
\newblock \bibinfo{journal}{\emph{CIKM}} (\bibinfo{year}{2020}).
\newblock


\bibitem[\protect\citeauthoryear{Zheng, Noroozi, and Yu}{Zheng
  et~al\mbox{.}}{2017}]%
        {zheng2017joint}
\bibfield{author}{\bibinfo{person}{Lei Zheng}, \bibinfo{person}{Vahid Noroozi},
  {and} \bibinfo{person}{Philip~S. Yu}.} \bibinfo{year}{2017}\natexlab{}.
\newblock \showarticletitle{Joint deep modeling of users and items using
  reviews for recommendation}. In \bibinfo{booktitle}{\emph{WSDM}}.
\newblock


\end{thebibliography}


\end{document}